\documentclass{article}      % onecolumn (second format)
\usepackage{ourmacros}

\begin{document}

\maketitle

% \begin{abstract}
%     \input{body/abstract}
% \end{abstract}

% \statekeywords

% \input{body/declarations}

% \pagebreak

% \section*{\centering \sc \thetitle}

\begin{abstract}
    The weighted Euler characteristic transform (\wect) is a new tool for extracting shape information from data equipped with a weight function.  
Image data may benefit from the \wect where the intensity of the pixels are used
to define the weight function.  In this work, an empirical assessment of the
\wect's ability to distinguish shapes on images with different pixel intensity
distributions is considered, along with visualization techniques to improve the
intuition and understanding of what is captured by the \wect.  Additionally, the
expected weighted Euler characteristic and the expected \wect are~derived.

\end{abstract}

% \statekeywords

\noindent
\textit{Funding} JCK and BTF acknowledge support from NSF under Grant Numbers
DMS 2038556 and 1854336. BTF and EQ also acknowledge support from NSF
under Grant Number 1664858.

\section{Introduction}\label{sec:intro}
Topological shape analysis seeks to summarize information about the shape of
data via topological invariants, which can then be used in subsequent analysis
tasks like classification or regression.  A classical topological invariant is the \emph{Euler Characteristic} (EC), which can easily be computed from data.
A popular extension of the EC for a filtered topological space is the
\emph{Euler characteristic function} (\fec),\footnote{
In some sources, this is referred to as the \emph{Euler characteristic curve},
but it is not a curve (as it is a step function).}
which tracks the EC as the filtration parameter changes.
When considering the shape of data embedded in $\R^d$, the shape is often filtered in different directions such as with the \emph{Euler Characteristic Transform} (ECT) and the \emph{Persistent Homology Transform} (PHT) \cite{turner2014persistent,ghrist2018persistent}.
These transforms have been used in a variety of settings such as
biology \cite{turner2014persistent,amezquita2022measuring}, oncology~\cite{crawford2020predicting}, and organoids \cite{marsh2022detecting}.
A recent generalization of the ECT that allows for a weighted simplicial complex
is the \emph{weighted Euler characterisitc transform}~(\wect)~\cite{jiang2020weighted}.
Jiang, Kurtek, and Needham~\cite{jiang2020weighted} extend work
of~\cite{turner2014persistent,ghrist2018persistent} and show that the \wect uniquely represents weighted simplicial complexes.
%We consider the case where we now have $k$ weight functions, one for each color channel.
The motivation of \cite{jiang2020weighted} is image data where the intensity of a pixel is used to assign the weights for the weighted simplicial complex.
We investigate the effectiveness of the \wect at discriminating shapes with
weights sampled from different distributions, and find the expected weighted EC
and the expected \wect of images under different weight distributions.

The effectiveness of the \wect at discriminating different types of images was
demonstrated in \cite{jiang2020weighted} where they considered the MNIST
handwritten digit dataset \cite{LeCun:1998aa} and magnetic resonance images of
malignant brain tumors (Glioblastoma Multiforme).  In the former example, the
\wect-based classification model outperformed classification models that used
either the image directly or the (unweighted) ECT in terms of ten-fold
cross-validation classification rate.
In this paper, we more generally explore the performance of the \wect when the
pixel intensities are sampled from different distributions.  In addition to
evaluating the performance of the \wect in a variety of settings, we improve the
interpretability of the \wect by estimating the \emph{expected} weighted EC and
the \emph{expected} \wect.

\section{The WECT}\label{sec:prelim}
In this paper, we use techniques from topological data analysis (TDA) to
study weighted shapes arising in image data. We write the
following definitions restricted to this setting, but note here that many of the
definitions can be stated more generally.  For a broader introduction to
TDA, see~\cite{edelsbrunner2010computational,dey2022computational}.

\subsection{From Images to Weighted Simplicial Complexes and
Filtrations}\label{sec:imgtowsc}

Computations of the \wect requires a representation of a weighted
simplicial complexes. In this paper, we compute the \wect using a simplicial
complex that represents an image.\footnote{For images, a cubical
complex provides a computationally preferable
representation~\cite{wagner2012efficient}; however, we choose
to use simplicial complexes in order to stay closer to the theory developed
in~\cite{jiang2020weighted}.}  A geometric~$k$-simplex $\sigma$ in
$\R^d$ is
the convex hull of~$k+1$ affinely independent points in $\R^d$.
For example, a zero-simplex is a point, a one-simplex is a segment, and a two-simplex is a triangle.
If we want to list the points of a simplex, we write $\sigma = [v_0,v_1, \ldots,
v_k]$ and, for each~$i \in \{ 0,1,\ldots, k\}$, we write $v_i \in \sigma$.
A simplex $\tau$ is a \emph{face} of
$\sigma$, denoted~$\tau < \sigma$, if it is the convex hull of a subset of those $k+1$ points.
A simplicial complex
is a collection of simplexes that are combined in
such a way that (i) for all simplexes~$\sigma \in K$, if $\tau$ is a face of
$\sigma$ then~$\tau \in K$, and (ii) if simplexes~$\sigma, \tau \in K$, then
either~$\sigma \cap \tau \in K$ or  $\sigma \cap \tau = \emptyset$.
$K_i$ is used to denote the set of $i$-simplices of~$K$.
The simplicial complex~$K$ is topologized using the Alexandroff topology~\cite{alexandroff1937diskrete}; with a slight abuse of
notation, $K$ denotes both the complex (as a set of simplices) and the underlying topological space.
\begin{figure}
    \centering
    \includegraphics[height=1.5in]{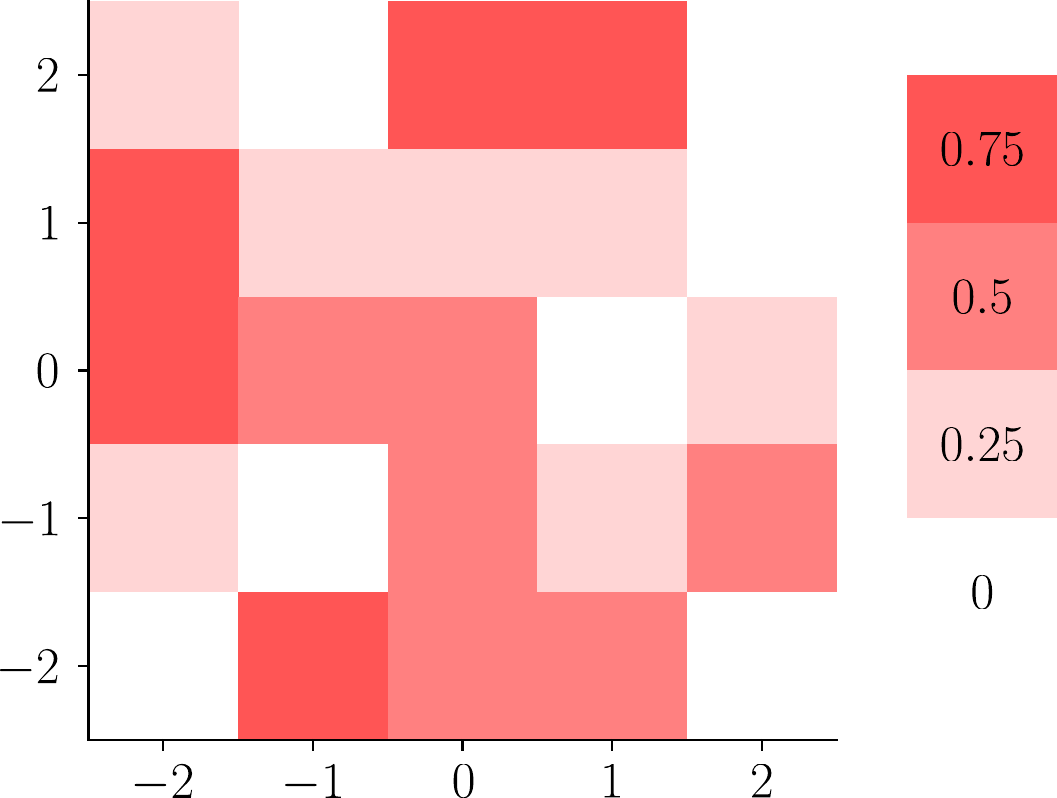}
    \caption{An image. We find \emph{shapes} in an image by thresholding the
    pixel intensities.}\label{fig:image-pixels}
\end{figure}

\paragraph{Shapes and Weights.}

\begin{figure}[tbh]
    \centering
    \begin{subfigure}{.23\textwidth}
        \centering
        \includegraphics[width=\linewidth]{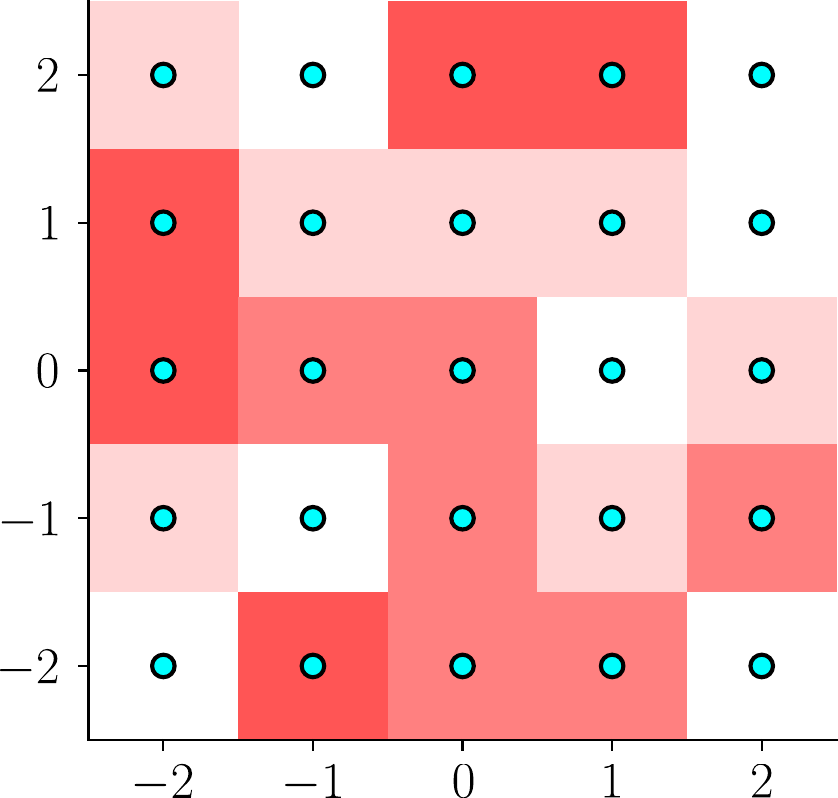}
        \caption{Vertices}
        \label{fig:image-pixvert}
    \end{subfigure}
    ~
    \begin{subfigure}{.23\textwidth}
        \centering
        \includegraphics[width=\linewidth]{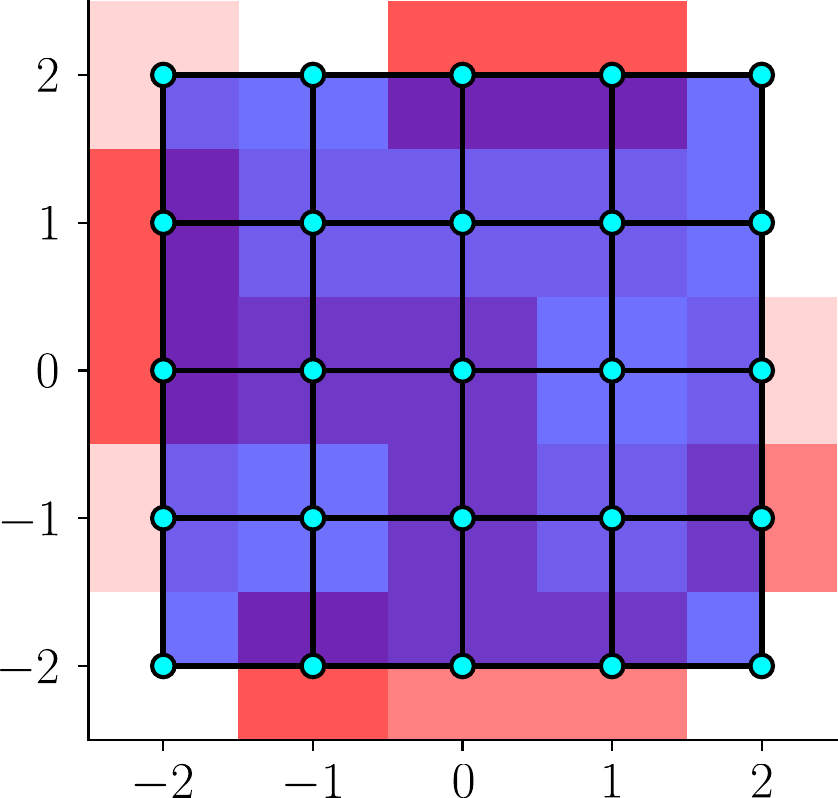}
        \caption{Cellulation}
        \label{fig:image-dual}
    \end{subfigure}
    ~
    \begin{subfigure}{.23\textwidth}
        \centering
        \includegraphics[width=\linewidth]{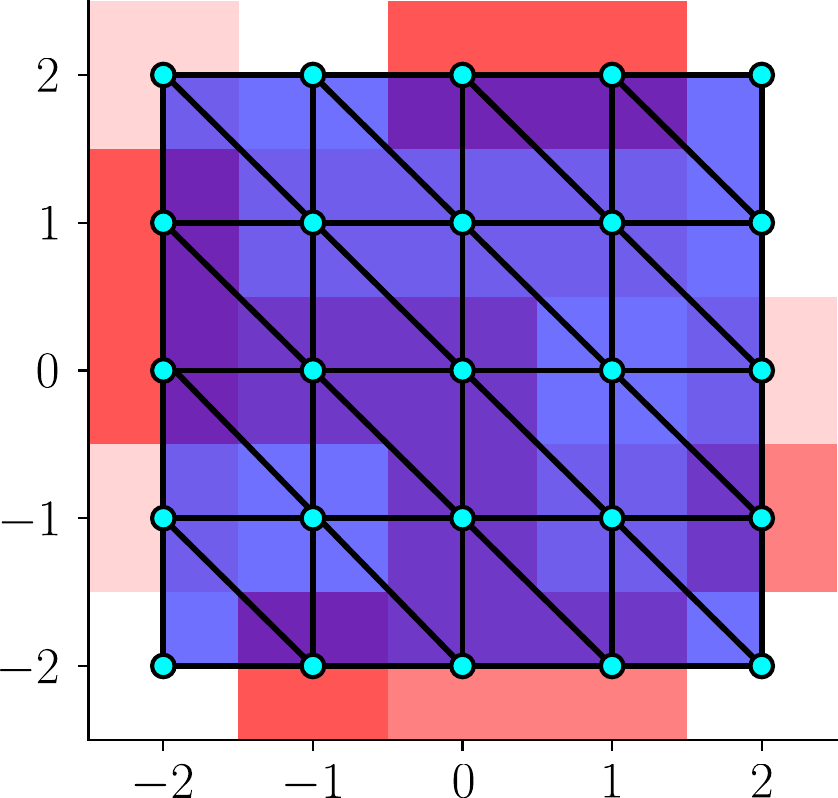}
        \caption{Triangulation}
        \label{fig:image-tri}
    \end{subfigure}
    ~
    \begin{subfigure}{.23\textwidth}
        \centering
        \includegraphics[width=\linewidth]{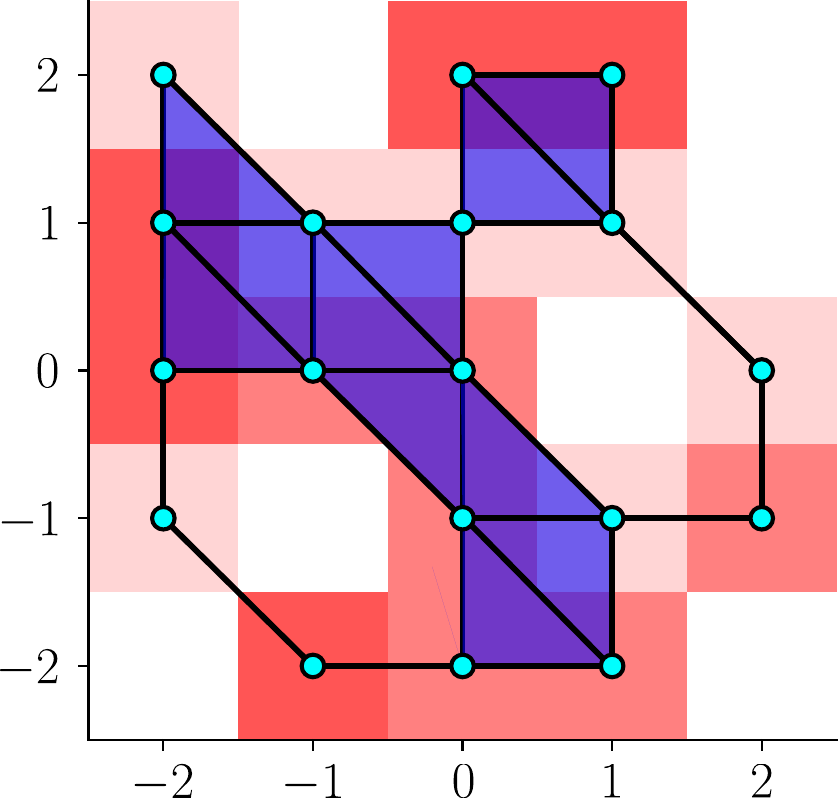}
        \caption{``Shape''}
        \label{fig:image-shape}
    \end{subfigure}
    \caption{
        Illustration of triangulating and thresholding an image.
        \insubfigref{image-pixvert}
        Vertices (zero-simplexes) are added to the center of each pixel.
        \insubfigref{image-dual}
        Edges (one-simplexes) are drawn between pixels that share an edge and
        squares added.
        \insubfigref{image-tri}
        A triangulation is created by drawing an edge between the top-left
        and bottom-right corners of each square.
        \insubfigref{image-shape}
        The simplicial complex resulting from restricting vertices to have positive intensity.
    }
    \label{fig:image}
\end{figure}

An image, such as the one shown in \figref{image-pixels} is a collection of
colored grid cells arranged in a regular grid.  The grid cells are often called
\emph{pixels}, and we consider the monochromatic setting (i.e., the pixel color
is represented by one number in $[0,1]$). Let $D$ be a rectangular subset of
$\Z^2$.
An image over $D$ is a function~$D \to [0,1]$.
The value assigned is called the \emph{pixel intensity}.
The points in~$\Z^2 \cap D$ represent the centers of the pixels.
Without loss of generality, we assume that $(0,0)$ is the center of the image;
% upper-left pixel and $(n-1,n-1)$ the center of the bottom-right pixel;
see
\subfigref{image}{pixvert}.
% \brittany{I think we want to assume $(0,0)$ is the
% center, right?} \jessi{We could assume this and the remove the centering step in the algorithm.}
%

In order to compute topological summaries of an image, we triangulate the image as follows.
Letting each pixel center be a vertex, an edge is added between
two vertices (pixels) that share a side, and a square is added when four pixels meet at a
corner.  A cellular structure is obtained that represents the continuous
$n \times n$ square domain~centered at $(0,0)$;
see \subfigref{image}{dual}.   Adding a diagonal from
the top-left to the bottom-right corners in each square and splitting each square
into two triangles, we obtain
the Freudenthal triangulation~\cite{freudenthal}; see
\subfigref{image}{tri}. The \emph{shapes} that we consider in this paper are
subcomplexes of this triangulation, often constructed by a threshold on the
pixel intensities, as shown in
\subfigref{image}{shape}.

Now that we have a simplicial complex, we explore how to extend the pixel
intensities to weights on each simplex.
In general, a \emph{weighted simplicial complex} is a pair:
$$ (K, \omega \colon K \to \R),$$
where $K$ is a simplicial complex, and $\omega$ is a function that assigns a
real value (i.e., a weight) to each
simplex in $K$.

Given a weight function $\omega \colon K_0 \to [0,1]$, we extend it to a
function over the whole complex,~$\Omega \colon K \to [0,1]$.
There are several options for performing this extension, including (i) the
\emph{maximum extension} ($\maxext$), which maps a~$k$-simplex
$\sigma_k=[v_0,v_1, \ldots, v_k]$ to the maximum value over its vertices,
(ii) the \emph{minimum extension} ($\minext$), which maps $\sigma_k$ to
the minimum value over its vertices, and (iii) the \emph{average extension}
($\avgext$), which maps $\sigma_k$ to the average value over its
vertices.  In other words, we have the following formulas:
\begin{align}
    \maxext(\sigma_k) &= \max_{i=0, \ldots, k} \omega(v_i), \label{eqn:max_extension} \\
    \minext(\sigma_k) &= \min_{i=0, \ldots, k} \omega(v_i), \text{ and }
    \label{eqn:min_extension} \\
    \avgext(\sigma_k) &= \frac{1}{k+1}\sum_{i=0}^k \omega(v_i). \label{eqn:avg_extension}
\end{align}

\begin{figure}
    \centering
    \begin{subfigure}{\textwidth}
        \centering
        \includegraphics[width=.8\textwidth]{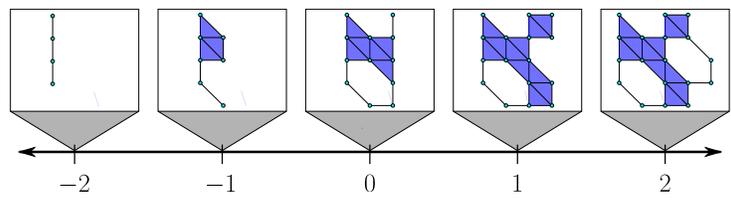}
        \caption{Filtration in direction $(1,0)$}
        \label{fig:filtered-right}
    \end{subfigure}
    \\
    \begin{subfigure}{\textwidth}
        \centering
        \includegraphics[width=.8\textwidth]{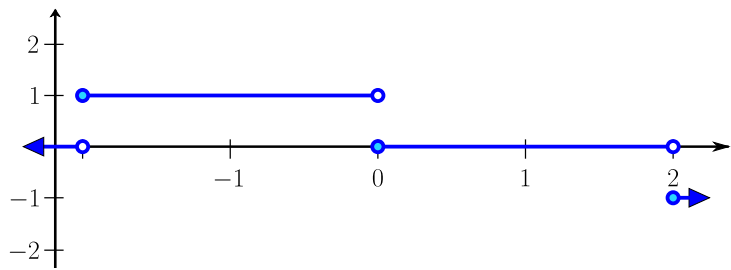}
        \caption{\ecf in direction $(1,0)$}
        \label{fig:iltered-ecf}
    \end{subfigure}
    \\
    \begin{subfigure}{\textwidth}
        \centering
        \includegraphics[width=.8\textwidth]{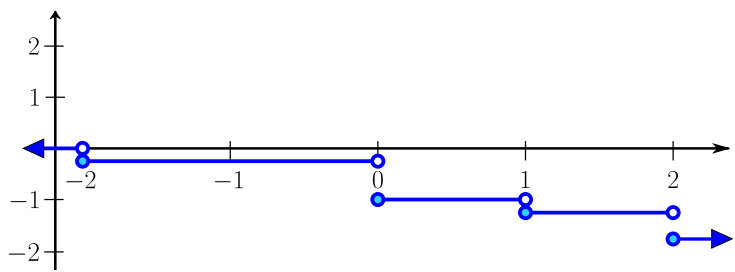}
        \caption{\wecf in direction $(1,0)$}
        \label{fig:iltered-wecf}
    \end{subfigure}
    \caption{Filtration of the shape from \subfigref{image}{shape}, along with
    the \ecf and \wecf. Note that there may be filtration values where \wecf
    changes, but \ecf does not (e.g.,   at $h=1$ above.
    }\label{fig:filtered}
\end{figure}

\paragraph{Filtrations.}
In what follows, we use lower-star filtrations parameterized by heights in
various directions.
The height of a point $v\in \R^d$ in direction~$s \in \sph^{d-1}$ is given by the
dot product $h_s(v):= v \cdot s$. Now, let $K$ be a simplicial complex in
$\R^d$. The height of a simplex $\sigma \in K$ is simply the maximum height
of any of its vertices; specifically, we have the $H_s \colon K \to \R$
defined by $H_s(\sigma) := \max_{v\in \sigma} v \cdot s$; we recognize this as the
maximum extension of $h_s$.  $H$ is an example of a \emph{filter function}, that
is a function on the simplices such that lower-level sets are always simplicial
complexes.
A \emph{filtration} is a parameterized family of topological spaces connected by
(forward) inclusions.
The \emph{lower-star filtration} is the following sequence of nested
simplicial~complexes:
\begin{equation}\label{eqn:lowerstar}
    \left\{ \bigcup_{t' \leq t} H_s^{-1}(-\infty,t'] \right\}_{t \in \Rbar}.
\end{equation}
For example, consider the shape from \subfigref{image}{shape}.
The lower-star filtration in direction $s=(1,0)$ is shown in
\subfigref{filtered}{right}.

\subsection{Euler Characteristic and Weighted Euler Characteristic Functions}

For a topological space $\X$, the EC of $\X$, denoted~$\chi(\X)$, is the alternating sum of Betti numbers,
$\chi(\X):=\sum_{i \in \N} (-1)^i \beta_i $.  If~$\X$ is a CW complex (in
particular, if it is a simplicial complex), $K$, then we have two alternate (but equivalent) definitions displayed below,
\begin{equation}\label{eqn:ec-altsum}
    \chi(K) = \sum_{i\in \N} (-1)^i |K_i| = \sum_{\sigma \in K} (-1)^{\dim(\sigma)}.
\end{equation}
The EC definition was generalized to include a weight function \cite{Betthauser:2018aa, jiang2020weighted}.
For a weighted simplicial complex $(K,\omega)$, the weighted EC may be defined as
\begin{equation}
\chi_{w}(K,\omega):=  \sum_{\sigma \in K}
(-1)^{\dim(\sigma)} \omega(\sigma).
\end{equation}
We call a function~$f \colon K \to \R$ a \emph{filter function} if for all $t
\in \R$, the sublevel set $f^{-1}(-\infty,t]$ is either empty or a simplicial
complex.
Given a filter function~$f \colon K \to \R$, we consider filtering $K$ by
$f$; that is, we have a nested family of sublevel sets~$f^{-1}(-\infty,t]$
for $t \in \R$. See \subfigref{filtered}{right} for a filtration of the shape in
\subfigref{image}{shape} in the direction $(1,0)$.

The EC of~$f^{-1}(-\infty,t]$ with respect to~$t$ is the
\emph{Euler characteristic
function~(\fec)},
denoted $\chi_{f} \colon \R \to
\Z$ and is defined~by
\begin{equation}
\chi_f(t) := \sum_{\sigma \in f^{-1}(-\infty,t]} (-1)^{\dim(\sigma)} = \chi (f^{-1}(-\infty,t]).
\end{equation}
If $f$ is a lower-star filter in direction $s$, then we may use
$s$ in the subscript instead of $f$.
Given a filter function~$f \colon K \to \R$ and an arbitrary function~$\Omega \colon K \to \R$, the \emph{weighted Euler
characteristic function (\fwec) for filter function $f$ with respect to the
weight function~$\Omega$} is defined as the function~$\chi_{f,\Omega} \colon
\R \to \R$, where
\begin{equation}
    \chi_{f,\Omega}(t) := \sum_{\sigma \in f^{-1}(-\infty,t]} (-1)^{\dim(\sigma)} \Omega(\sigma).
\end{equation}
We follow the setting of~\cite{jiang2020weighted}, where $f$ is a lower-star filter (i.e., a discretized
version of a height filter) in a given direction in $\sph^{d-1}$; see \eqnref{lowerstar}.
Again, if $f$ is a lower-star filter in direction $s$, then we may use
$s$ in the subscript instead of~$f$.
See \figref{filtered} for an example \ecf and \wecf for a filtered shape.

\subsection{Weighted Euler Characteristic Transform}

Using an ensemble of directional filter functions, one for
each direction in~$s \in \sph^{1}$, a parameterized
collection of \fecs, the ECT, is defined as the
function~$\text{ECT}_K \colon \sph^1 \times \R \to \Z$, where\footnote{The ECT and the
\wect can be generalized to higher dimensions, but our setting is focused on 2D
images so here we consider directions in $\sph^{1}$ only.}
\begin{equation}
    \text{ECT}_K(s,t) = \chi_{s}(t).
\end{equation}
For each direction $s \in \sph^1$, we can think of this as a function $\R \to
\Z$, namely~$\chi_{s}$ described above.
No two distinct shapes have the same ECT~\cite{turner2014persistent}.
Since its introduction, the ECT has developed both in theory and in
practice~\cite{ghrist2018persistent,jiang2020weighted,berry2020functional,marsh2022detecting,crawford2020predicting}.

As noted previously, the \wect is a generalization of the ECT for weighted
simplicial complexes, and was introduced in \cite{jiang2020weighted}. Let $K$ be a
simplicial complex in $\R^2$ and let $\Omega \colon K \to \R$ be its weight
function (note that~$\Omega$ need not be a filter function).
The \wect for $K$ assigns to each~$s \in \sph^1$ the
function~$\text{WECT}_{K,\Omega}(s,t) \colon \R \to \Z$ defined~by:
\begin{equation}
    \text{WECT}_{K,\Omega}(s,t) = \sum_{d=0}^{2}(-1)^{d}\sum_{\substack{\sigma \in K_{d} \\
                    H_s(\sigma) \leq t}} \Omega(\sigma)
        = \sum_{\substack{\sigma \in K \\ H_s(\sigma) \leq
                    t}}(-1)\textsuperscript{dim($\sigma)$}\Omega(\sigma),
\end{equation}
where we recall that~$H_s \colon K \to \R$ is the lower-star filter in direction $s$.
When subscripts are difficult to read, we
use~$\text{WECT}(s,t;K,\Omega):= \text{WECT}_{K,\Omega}(s,t)$.

\paragraph{Distances Between \wects.} \label{par:distances}

Distances between \wects is necessary for analysis tasks such as classification or inference.
To define a distance in our setting, consider two weight functions $\omega\colon
K \to \R$ and $\omega' \colon K' \to \R$, with two \wects for each
direction $s \in S^{d-1}$ denoted as $W_s \colon \R \to \N$ and
$W'_s \colon \R \to \N$, respectively.
Since $W_s,W_s'$ are $\N$-valued
functions, different distance metric between functions may be considered to compare them.
Let $\rho \colon
\mathcal{F} \times \mathcal{F} \to \R$ be that distance metric, where
$\mathcal{F}$ is the space of all functions $\R \to \N$.  Common metrics are
induced from the sup norm or the $L^p$ norm.\footnote{The distance proposed in~\cite{jiang2020weighted} is the standard $L^2$ norm.}
 Then, we define the distances
between two \wects, $W = \{W_s\}_{s\in \sph^1}$ and $W'= \{W'_s\}_{s\in \sph^1}$, as $D \colon \mathcal{W} \times \mathcal{W} \to \R$ by integrating the following:
\begin{equation}\label{eqn:wect-dist}
    D(W, W') := \int_{s \in S^{d-1}} \! \rho(W_s, W_s')  \, \mathrm{d}s.
\end{equation}

\subsection{Algorithm and Vectorizing \wects}

In \algref{wect}, we provide the algorithm used to compute the \wect of a given
weighted shape image.
We recall that the \wect is the family of \wecfs, one for each direction
in~$\sph^1$. Computationally, knowing all \wecfs is infeasible so often much fewer are used.   We first consider how to compute the \wecf in one direction, $s$,
as given in \alglnsref{wect-start-s}{wect-end-s} for $(K,\Omega)$.
\begin{algorithm}[tbh]
    \caption{$\textsc{Get-WECT}(S,K,\ell, \Omega)$}
    \label{alg:wect}
    \begin{algorithmic}[1]
        \REQUIRE Directions $S \subset \mathbb{S}^{1}$;
            triangulation~$K$;
            vertex locations $\ell \colon K_0 \to \R^2$;
            weight function~$\Omega \colon K \to \R $
        \ENSURE vectorized weighted Euler characteristic transform of $(K,\Omega)$

        \FOR{$s \in S$}
            \STATE Define height function $h_s\colon K_0 \to \R$ by $h_s(v)=s
                \cdot \ell(v)$\label{algln:wect-start-s}
            \STATE Define $H_s \colon K \to \R$ as the maximum extension
                of~$h_s$\label{algln:wect-filter}
            \STATE $\{\sigma_1,\sigma_2, \ldots, \sigma_n\} \gets$ ordering
                simplices in $K$ by $H_s$ (ties
                broken arbitrarily)\label{algln:wect-ordersimps}
            \STATE $i \gets 0$
            \STATE $curheight \gets -\infty$ \COMMENT{1.65in}{current height being processed}
            \STATE $\chi \gets 0$ \COMMENT{2.4in}{running Euler characteristic}
            \STATE $WECT_s \gets \emptyset$
            \WHILE{$i < n$}
                \STATE $i++$
                \IF{$H_s(\sigma_i) \neq curheight$}
                    \STATE Append $(curheight, \chi)$ to
                    $WECT_s$\label{algln:wect-addpiece}
                    \STATE $curheight \gets H_s(\sigma_i)$
                \ENDIF
                \STATE $\chi += (-1)^{\dim(\sigma_i)} \Omega(\sigma_i)$
            \ENDWHILE
            \STATE $W_s \gets$ vectorization of $WECT_s$ \label{algln:wect-end-s}
        \ENDFOR

        \RETURN Concatenation of $W_s$ for all~$s \in S$\label{algln:wect-concat}

    \end{algorithmic}
\end{algorithm}

\paragraph{Computing a Weighted Euler Characteristic.} The \wecf is a step
function.  To represent it (in the variable $WECT_s$ of \algref{wect}), we use
tuples $(a,b)$, which is read as ``the weighted Euler characteristic of $H_s^{-1}(-\infty,t]$ for all
$t$ starting at $a$ until the next tuple is~$b$.''  By storing the two-tuples, we have an exact representation of the \wecf.

To compute the \wecf of $(K,\Omega)$ in direction $s$, we first compute the
filter function ($H_s$; defined in \alglnref{wect-filter}), then order the
simplices in non-decreasing function order (\alglnref{wect-ordersimps}).
Considering these simplices in order, we keep a running tally of the EC, and add
a two-tuple to $WECT_s$ each time a
new height is encountered (\alglnref{wect-addpiece}).  In the end, we have a
list of two-tuples~$\{ (h_i, X_i)\}_{i\in \{1, 2, \ldots, N\}}$, where $h_1 <
h_2 < \cdots < h_N$.  Then, the \wect for direction $s$ is:
\begin{equation}
    \text{WECT}(s,t;K,\Omega) =
    \begin{cases}
        0 & t < 0 \\
        X_i & t \in [h_i,h_{i+1}] \\
        X_N & t \geq h_N.
    \end{cases}
\end{equation}

\paragraph{Vectorizing the \wect.}
The exact representation of the \wecf, as a piecewise defined function, is challenging to
work with for tasks like classification where alignment across different \wecfs
is needed.  So, we choose a vectorization parameter, $n_v$, and
we vectorize by sampling~$n_v$ uniformly-spaced values
over an interval of heights that is context-specific.

We also must choose the directions.  One approach is to choose the number of sample
directions, $n_s$, then select directions whose angles with the positive $x$ axis
are: $\{ 0,
\frac{2\pi}{n_s}, \frac{2\cdot 2\pi}{n_s}, \ldots,  2\pi - \frac{1}{n_s}\}$.
Thus, we have $n_s$ equally-spaced direction vectors.\footnote{More generally, we can choose an $\epsilon>0$ and use an $\epsilon$-net~\cite{}.}

To put this all together, in \alglnref{wect-concat}, we concatenate the
\wects for the different directions in order of increasing angle with the
positive $x$ axis.

\paragraph{Relating Vectorization to Exact \wect.}
We define a distance between the vectorized representations such that
as the vectorization and sampling variables increases ($n_v \to \infty$ and $n_s \to \infty$,
respectively), we approach the distance given in \eqnref{wect-dist}.  Letting
$\rho$ denote the two-norm, we have:
\begin{equation}\label{eqn:vect-wect-dist}
    \hat{D}(W, W') := \sum_{s \in S} \! ||WECT_s - WECT_s'||_2.
\end{equation}
As a result, if $n_s$ and $n_v$ are large enough, we can perform tasks such as
clustering on these vectors and approximate working in the original function
space itself.

\paragraph{Image Data.}

In this paper, $K$ represents a shape in an $n \times n$ image $A$.
We describe the input to \algref{wect} for image data:
\begin{itemize}
    \item $S$: Select $n_s$, the number of sampled directions, then starting with
        the first direction $(0,1) \in \sph^{1}$, select $n_s$ equally-spaced directions over
        $\sph^1$.
    \item $K$: We use the process described in \secref{imgtowsc}
        to find a subcomplex of the Freudenthal
        triangulation representing the ``shape'' in the image.
    \item $\ell$: Throughout this paper, we assume that $n$ is odd.  The $n
        \times n$ grid of vertex locations (corresponding to centers of pixels)
        are the integer coordinates in the square
        $[-\frac{n-1}{2},\frac{n-1}{2}]^2$.
    \item $\Omega$: Let $\omega$ be the function that takes as input a pixel
        (vertex of $K_0$) and returns the corresponding value of $A$.  Then,
        $\Omega$, is one of the extensions of~$\omega$ described in
        \secref{imgtowsc}.
\end{itemize}

In the output, we use the parameter $n_s$ for the number of samples used to
vectorize each directional \wecf.  These values are uniformly sampled over the
interval $(-\frac{n-1}{2},\frac{n-1}{2})$ if $n$ is odd.

\subsection{Theoretical Results on Expected Weighted Euler Characteristic}

In \secref{imgtowsc}, we discussed several ways to extend a weight function on
the vertices to a weighted simplicial complex.
Here, we investigate how the
choice of extension affects the weighted EC.  In what follows, let $K$ be a
two-complex, and let $\omega \colon K_0 \to \R$ be the assignment of intensities
to the vertices, drawn independently from a distribution
centered at $0.5$.

\subsubsection{Average Extension}

Let $\avgext \colon K \to \R$ be the average extension of $\omega$.
For each vertex $v \in K_0$, since the pixel intensity is drawn from a
distribution centered at $0.5$, we know that~$\E(\avgext(v))=\E(\omega(v))=0.5$.
In the average extension, we recall that the value of an edge and a triangle are
\begin{eqnarray}
    \avgext( [v_0,v_1]) &= &\frac{1}{2} \left( \omega(v_0)+\omega(v_1)
        \right)\\
    \avgext ( [v_0,v_1,v_2]) &= &
        \frac{1}{3} \left( \omega(v_0)+\omega(v_1) +\omega(v_3) \right),
\end{eqnarray}
respectively.
And so, by the linearity of expectation, we compute the expectation of the
function value on edges and triangles:
\begin{equation}
    \E\left(\avgext([v_0,v_1])\right)=\frac{1}{2}
    (\E(\avgext(v_0))+\E(\avgext(v_1))) = \frac{1}{2} (0.5+0.5)=0.5
\end{equation}
and
\begin{eqnarray}
    \E(\avgext([v_0,v_1,v_2]))
        &=& \frac{1}{3} (\E(\avgext(v_0)) +\E(\avgext(v_1))+\E(\avgext(v_2)))\\
        &=& \frac{1}{3}(0.5+0.5+0.5)=0.5.
\end{eqnarray}
Thus, for any $\sigma \in K$, we have shown that~$\E(\avgext(\sigma))=0.5$, regardless of
the distribution from which the pixel intensities were drawn (as long as that
distribution is centered at $0.5$).

To compute the expected weighted EC, we again use linearity of
expectation to find:
\begin{align}
    \E\left(\chi_{w}(K,\avgext)\right)
        &=  \E \left( \sum_{\sigma \in K}(-1)^{\dim(\sigma)}\avgext(\sigma) \right) \\
        &=  \sum_{\sigma \in K}(-1)^{\dim(\sigma)} \E(\avgext(\sigma)) \\
        &=  \frac{1}{2}\sum_{\sigma \in K}(-1)^{\dim(\sigma)}  \\
        &= \frac{1}{2} \chi(K)\label{eqn:weightedtochi}
\end{align}
Thus, the expected weighted EC relies (not surprisingly) on computing the EC,
which, by Equation~\eqnref{ec-altsum}, we know if we have the number of simplices of each
dimension.

\begin{example}{Square Image with Average Extension}\label{ex:square-avg}
    Consider an $n \times n$ random image, where the pixels of the image
    are drawn independently from a distribution
    centered at $0.5$.
    Let~$K$ be the corresponding Freudenthal triangulation.
    Let~$\omega \colon K_0 \to [0,1]$ be this assignment of values
    to the vertices/pixels.
    Counting the simplices, we have~$n^2$ vertices (one centered in
    each pixel),~$(n-1)n$ horizontal edges,~$(n-1)n$ vertical edges, $(n-1)^2$
    diagonal edges, and $2(n-1)^2$ triangles.  And so, the EC, $\chi(K)$, is:
    \begin{align}
        \chi(K)
        &= n^2-2(n-1)n-(n-1)^2+2(n-1)^2 \\
        &= (3n^2-4n+2)-(3n^2-4n+1)\\
        &= 1.
    \end{align}
    Plugging this into \eqnref{weightedtochi}, we find that the expected weighted
    EC for a square image is $0.5$, regardless of the size of the image.
    In fact, any random image whose support has trivial homology (i.e., no holes in
    the image), has the same expected weighted EC.
    Note that this only assumes that all the pixels are drawn independently from a
    distribution centered at~$0.5$.
\end{example}

The shapes that we are interested in are
subcomplexes of the $n \times n$ image obtained by thresholding on intensities.
Thus, a complex may have some pixels not included (and hence any edge/triangle
they are incident to not included), resulting in different
ECs.  Therefore, the corresponding expected weighted EC would
be different for different shapes.

\subsubsection{Maximum and Minimum Extension}

Unlike the average extension, the expected value of the weighted EC for the
maximum and minimum extensions are dependent on the distribution of the pixel
intensities.  Assuming the pixel intensities are drawn from a
$U(0,1)$, the
distribution of the $k$th order statistic,~$Y_{(k)}\sim$Beta$(k,
n+1-k)$, where
$n$ is the number of vertex intensities considered.\footnote{For more details
about distributions of order statistics under different underlying distributions
(beyond the uniform distribution), see \cite{david2004order}.}  In particular,~$Y_{(1)}$ is the minimum and $Y_{(n)}$ is the maximum.
The expected value of $Y_{(k)}$ is~$\frac{k}{n+1}$.
Therefore, the expectation of the function value on edges and triangles for the maximum extensions are:
\begin{equation}
    \E\left(\maxext([v_0,v_1])\right)= \E\left(\max(\omega(v_0), \omega(v_1))\right) = \frac{2}{3},
\end{equation}
and
\begin{equation}
    \E(\maxext([v_0,v_1,v_2]))= \E\left(\max(\omega(v_0), \omega(v_1), \omega(v_2))\right) = \frac{3}{4}.
\end{equation}
(Using the minimum extension, we also have the $\E\left(\minext([v_0,v_1])\right) =
\frac{1}{3}$ and the  $\E(\minext([v_0,v_1,v_2]))=\frac{1}{4}$.)
Then, the expected value of the weighted EC is
\begin{align}
    \E\left(\chi_{w}(K,\maxext)\right) &=  \E \left( \sum_{\sigma \in K}(-1)^{\dim(\sigma)}\maxext(\sigma) \right) \\
        &=  \sum_{\sigma \in K}(-1)^{\dim(\sigma)} \E(\maxext(\sigma)) \\
        &=  \sum_{i \in \N}\sum_{\sigma \in K_i}(-1)^{\dim(\sigma)} \left(\frac{i+1}{i+2}\right) \\
        &= \sum_{i \in \N} (-1)^i \left(\frac{i+1}{i+2}\right) |K_i|.
\end{align}
We return to the example of the $n \times n$ image.

\begin{example}{Square and Rectangular Images with Maximum and Minimum
    Extensions}\label{ex:square-max}
    Considering the same example of an $n\times n$ image from
    \exref{square-avg}, the expected value of the weighted EC using the maximum extension is
    \begin{align}
    \E(\chi_{w}(K,\maxext)) &=  \left(\frac{1}{2}\right) n^2 -\left(\frac{2}{3}\right) (3n^2-4n+1) + \left(\frac{3}{4}\right) (2)(n-1)^2 \\
        &= \frac{5-2n}{6}.
    \end{align}

    For the minimum extension, $\E(\chi_{w}(K,\minext))=\frac{2n+1}{6}$. More
    generally, the expected weighted EC for an $n \times m$
    image using the maximum extension is~$\frac{5-n-m}{6}$ and using the minimum
    extension is $\frac{n+m+1}{6}$.
    Notice that if $n=1$, both~$\E(\chi_{w}(K,\maxext))=\E(\chi_{w}(K,\minext))=1/2$ because there is only one vertex and no higher order simplexes.
\end{example}

\subsubsection{Expected \wect}

The expected \wect can be derived by looking at each direction independently and
extending the ideas from above.  For
example, if the pixel intensity distribution is centered at $0.5$ and $\Omega
\colon K \to \R$ is the weight function, then the expected WECF for direction $s \in \sph^1$ is a
function $\R \to \R$ that takes $t \in \R$ and maps it to
$\E\left( \text{WECT}_K(s,t, \avgext) \right)$.

In particular, for the average extension, we have:
\begin{align}
    \E\left( \text{WECT}_K(s,t, \avgext) \right)
        &=\E\left(\chi_{s,\avgext} (t)\right)\\
        &= \sum_{\substack{\sigma \in K \\ H_s(\sigma) \leq
                    t}} (-1)^{\dim(\sigma)} \E\left(\avgext(\sigma)\right) \\
        &= \frac{1}{2} \sum_{\substack{\sigma \in K \\ H_s(\sigma)
                    \leq t}}(-1)^{\dim(\sigma)} \\
        &= \frac{1}{2} \chi(\avgext^{-1}(-\infty,t]),\label{eqn:expWECTSquare}
\end{align}
where we recall that~$H_s \colon K \to \R$ is the lower-star filter in direction $s$.

For the maximum extension, we have:
\begin{align}
    \E\left( \text{WECT}_K(s,t, \maxext) \right)
        &=\E\left(\chi_{s,\maxext} (t)\right)\\
        &= \sum_{\substack{\sigma \in K \\ H_s(\sigma) \leq
                    t}} (-1)^{\dim(\sigma)} \E\left(\omega(\sigma)\right) \\
        &= \sum_{i \in \N} \sum_{\substack{\sigma \in K_i \\ H_s(\sigma) \leq
                    t}} (-1)^{\dim(\sigma)} \left(\frac{i+1}{i+2}\right)  \\
        &= \sum_{i \in \N} \left(\frac{i+1}{i+2}\right) \sum_{\substack{\sigma \in K_i \\ H_s(\sigma) \leq
                    t}} (-1)^{i}\\
        &= \sum_{i \in \N} \left(\frac{i+1}{i+2}\right) (-1)^i |K_i(t)|,
\end{align}
where we use $K_i(t)$ to denote the set of $i$-simplices $\sigma \in K$ such that
$H_s(\sigma) \leq t$.

\begin{example}{Square Image}
    Extending the example of an $n\times n$ image with the maximum
    extension from
    \exref{square-max}. Let $s \in \sph^1$ and $\Omega=\avgext$.
    Let~$h_0=\min_{\sigma \in K} \Omega(\sigma)$.  Then, the expected \wecf in direction $s$ is the function $\R
    \to \R$ defined by
    \begin{equation}
        t \mapsto
            \begin{cases}
                0 & t <h_0 \\
                0.5 & t \geq h_0
            \end{cases}
    \end{equation}

    Now, consider $s=(0,1)$, $\Omega=\maxext$, and $n$ odd.  Note that, the
    range of the height function~$H_s$
    is~$\{-{\frac{n-1}{2}}, -{\frac{n-1}{2}}-1, \ldots, -1,0,1,
    \ldots,{\frac{n-1}{2}}\}$.
    Letting $m_t$ denote the number of columns in the image (that is,
    $m_t=\ceil{t}+\frac{n-1}{2}$),
    the expected \wecf in direction
    $s$ is the function $\R \to \R$ defined by
    \begin{equation}
        t \mapsto
            \begin{cases}
                0 & t < -\frac{n-1}{2} \\
                \frac{1}{6}(5-n-m_t) & t \in (-\frac{n-1}{2},\frac{n-1}{2})\\
                \frac{5-2n}{6} & t \geq \frac{n-1}{2}.
            \end{cases}
    \end{equation}
    A simple check shows that this equation is continuous.
    The middle equation comes from the fact
    that $K(t):= \{\sigma \in K ~|~ H_s(\sigma) \leq t\}$ is a rectangular
    image that is $n$ pixels tall and $m_t=\ceil{t}-\frac{n-1}{2}$
    pixels wide.
\end{example}

\section{Empirical Study}\label{sec:spaces}
In this section, we assess the performance of using \wects to represent images generated under
different weight distributions for the pixel intensities, different domain shapes,
and different function extensions for assigning weights to the simplexes.
We then build classification models for each pair of image types in order to
evaluate the informativeness of \wects.
Rather than seeking to minimize the misclassification rate,
the goal is to understand how changing intensity distributions affects the classification performance.
% The performance of \wects is compared to the performance using the same classification model (i.e., support vector machines), but with different inputs, namely, (unweighted) ECTs and the original image \jessi{Assuming we do these comparsons}.
% \jessi{We also compute the estimated \wect for each of the images settings investigated.}

%

\subsection{Data Sets}
We generated image datasets
using the following parameters: shape, pixel intensity distribution,
and function extension (see Equations~\eqref{eqn:max_extension}-\eqref{eqn:avg_extension}).
The domain of an image is a $65\times 65$ grid centered at $(0,0)$,
with vertices on an integer lattice, triangulated using the Freudenthal triangulation~\cite{freudenthal}; see
\subfigref{image}{tri}.

\begin{figure}
    \centering
    \begin{subfigure}{.23\textwidth}
        \centering
       \includegraphics[width=\linewidth]{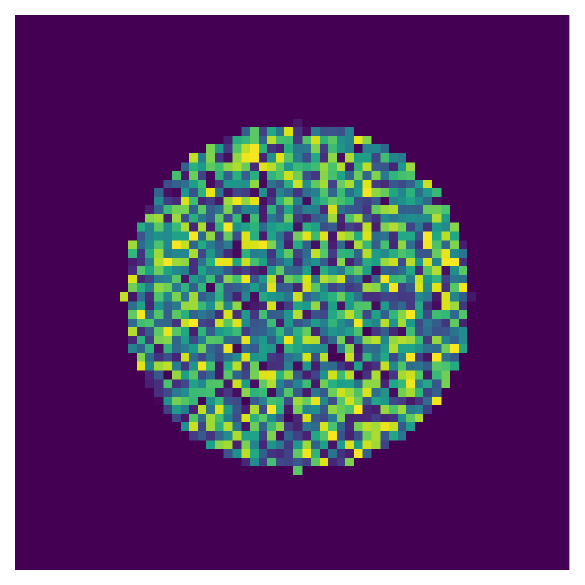}
        \caption{Disc}
        \label{fig:disc}
    \end{subfigure}%
    \hfill
    \begin{subfigure}{.23\textwidth}
        \centering
       \includegraphics[width=\linewidth]{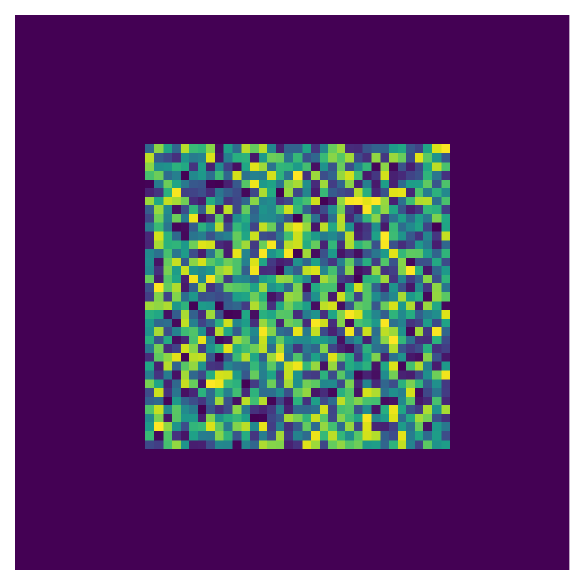}
        \caption{Square}
        \label{fig:square}
    \end{subfigure}%
    \hfill
    \begin{subfigure}{.23\textwidth}
        \centering
       \includegraphics[width=\linewidth]{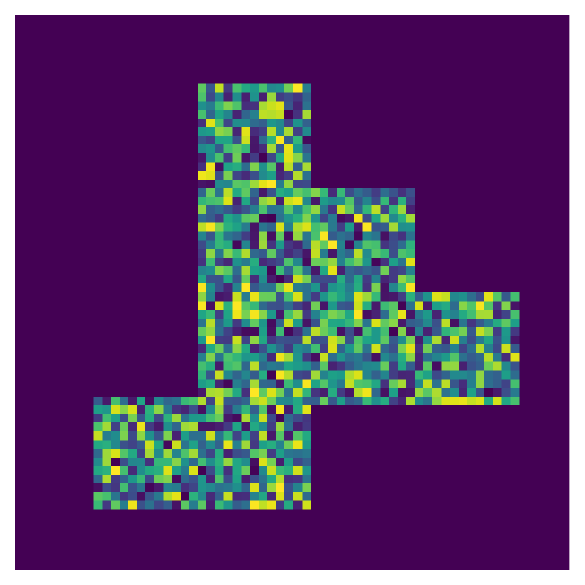}
        \caption{Tetris}
        \label{fig:tetris}
    \end{subfigure}
    \hfill
    \begin{subfigure}{.23\textwidth}
        \centering
        \includegraphics[height=1.1in]{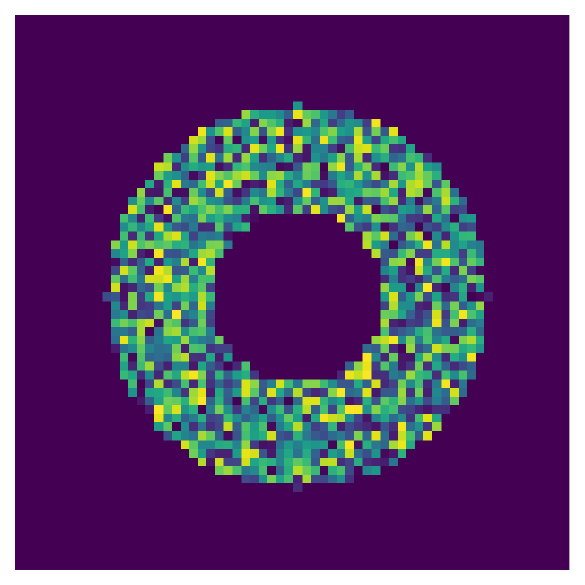}
        \caption{Annulus}
        \label{fig:colorbar}
    \end{subfigure}

    \vspace{0.5em} % Add vertical spacing between rows

    \begin{subfigure}{.23\textwidth}
        \centering
       \includegraphics[width=\linewidth]{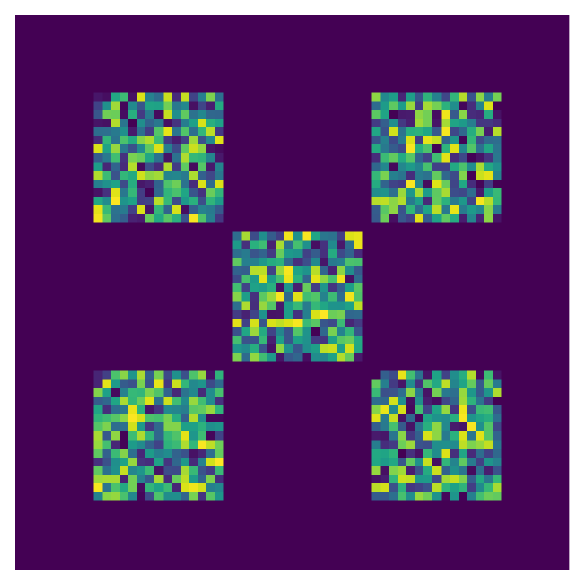}
        \caption{Clusters}
        \label{fig:clusters}
    \end{subfigure}%
    \hfill
    \begin{subfigure}{.23\textwidth}
        \centering
       \includegraphics[width=\linewidth]{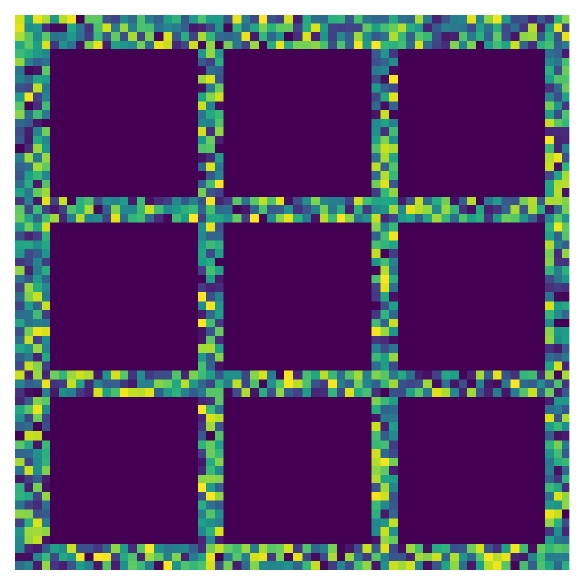}
        \caption{Swiss cheese}
        \label{fig:swiss}
    \end{subfigure}%
    \hfill
    \begin{subfigure}{.23\textwidth}
        \centering
        \includegraphics[width=\linewidth]{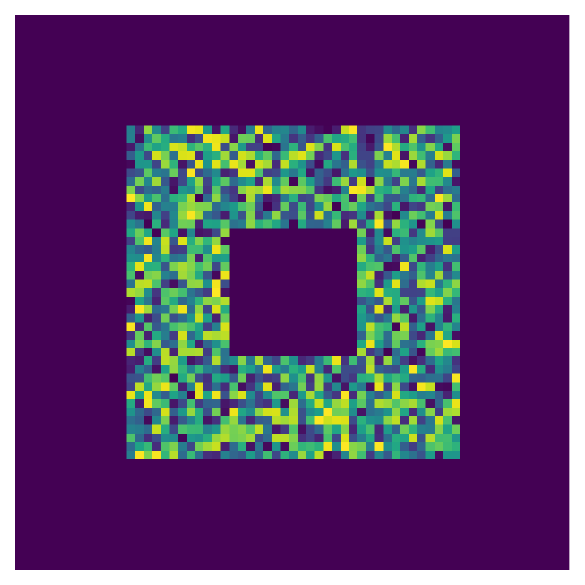}
        \caption{Square annulus}
        \label{fig:square_annulus}
    \end{subfigure}
    \hfill
    \begin{subfigure}{.23\textwidth}
        \centering
        \includegraphics[height=1.1in]{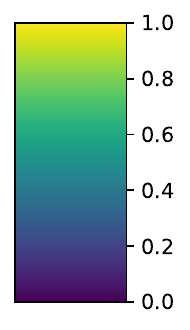}
        \caption{Intensity scale}
        \label{fig:colorbar}
    \end{subfigure}
    \caption{Examples of image shapes. The simulation study includes the following shapes illustrated here:
    (a) disc, (b) square, (c) Tetris, (d) annulus, (e) clusters, (f) Swiss cheese, and (g) square annulus.
    The intensity weights were drawn from a $U(0,1)$ distribution, but three normal distributions are also considered in the study.
    The intensity scale is shown in subfigure (h).}
    \label{fig:shapes}
\end{figure}
\paragraph{Shape.}
The \emph{shapes} are the domain of the image pixels with positive intensity.
We consider the following shapes:  disc, annulus, square, Tetris, clusters, and Swiss
cheese; see \figref{shapes} for example images of these shapes.
% \camera{should we not have distribution showing here too? - BTF}
%
When creating these shapes, we ensure that
they have approximately the same number of non-zero pixel intensities  (of the
$4225$ pixels in the $65 \times 65$ grid); see
\tabref{nonzeroPixels}.
\begin{table}[]
    \centering
    \caption{Number of nonzero pixels for each shape in the $65\times65$ grid.}
    \label{tab:nonzeroPixels}
    \begin{tabular}{r|llll}
        \hline
        Shape & Nonzero Pixels \\
        \hline
        Disc & $1257$ \\
        Square & $1225$ \\
        Tetris & $1249$ \\
        Annulus & $1212$ \\
        Clusters & $1125$ \\
        Swiss Cheese & $1495$ \\
        Square Annulus & $1296$ \\
    \end{tabular}
\end{table}

\begin{figure}[tbh]
\centering
\includegraphics[height=2.25in]{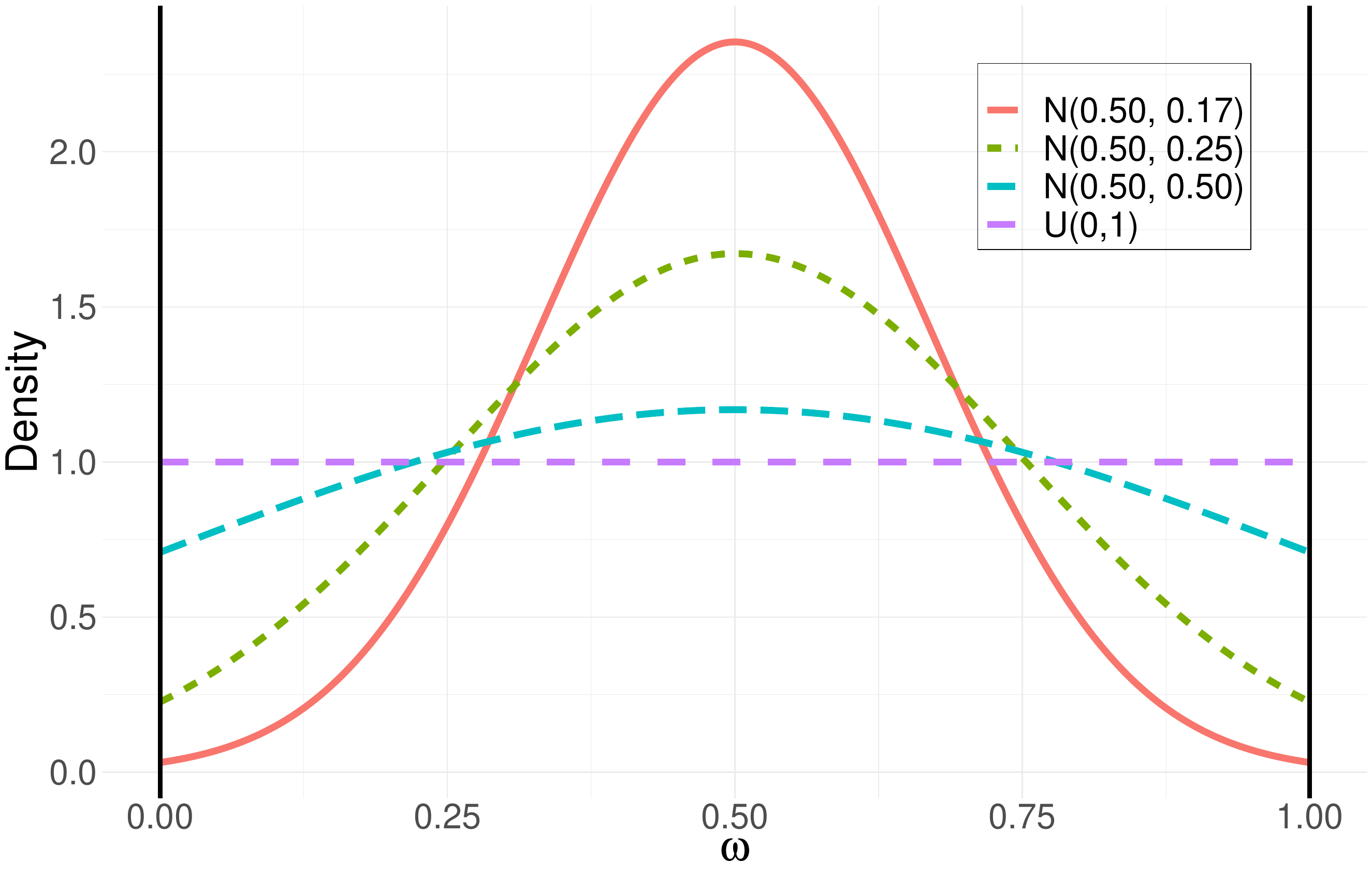}
\caption{The probability densities for the four pixel intensity distributions are displayed.  The normal distributions are truncated to $(0,1)$.}\label{fig:pixel_distributions}
\end{figure}
\begin{figure}
    \centering
    \begin{subfigure}{.47\textwidth}
        \centering
       \includegraphics[height=1.5in]{figs/shapes/annulus_uniform.pdf}
        \caption{weight$\sim U(0,1)$}
        \label{fig:image-uniform_weights}
    \end{subfigure}
    ~
    \begin{subfigure}{.47\textwidth}
        \centering
       \includegraphics[height=1.5in]{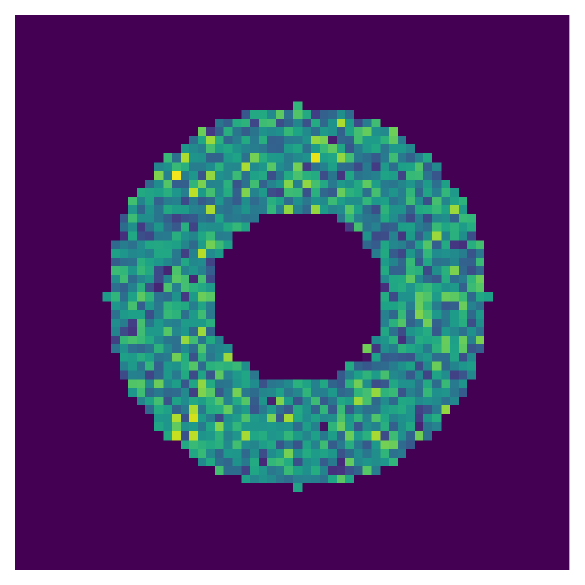}
        \caption{weight$\sim N[0.5, 0.17]$}
        \label{fig:image-normal_weights}
    \end{subfigure} \\
    % %
    %     \begin{subfigure}{.47\textwidth}
    %     \centering
    %    \includegraphics[width=\linewidth]{figs/uniform_annulus_wect_max.pdf}
    %     \caption{\wect for $U(0,1)$}
    %     \label{fig:image-uniform_weights_wect}
    % \end{subfigure}
    % ~
    % \begin{subfigure}{.47\textwidth}
    %     \centering
    %    \includegraphics[width=\linewidth]{figs/n17_annulus_wect_max.pdf}
    %     \caption{\wect for $N(0.5, 0.17)$}
    %     \label{fig:image-normal_weights_wect}
    % \end{subfigure}
    %
    \caption{An example image of an annulus with pixel intensities drawn from
    the distributions
    \insubfigref{image-uniform_weights}~$U(0,1)$ and
    \insubfigref{image-uniform_weights}~$N(0.5, 0.17)$.
    % (c) and (d) show the corresponding \wects of the images in (a) and (b) for eight directions.
    }
    \label{fig:image-annulus}
\end{figure}

\paragraph{Pixel Intensity.}
We consider four weight distributions for the pixel intensities:  a uniform and three truncated normal distribution with
different standard deviations.
The uniform distribution takes values greater in the interval~$(0,1]$, and is
denoted~$U(0,1)$.
The truncated normal is centered at $\mu=0.5$ with a standard deviation of
$\sigma$, $N(0.5, \sigma)$, and truncated to take values in the interval~$(0,1]$.
The $\sigma$ takes three values so that $\mu \pm i \sigma$, for~$i=1,2,3$, is
(approximately) the support, $(0,1]$; in other words, we consider the
distributions~$N(0.5, 0.17)$, $N(0.5, 0.25)$, and $N(0.5, 0.5)$.
\figref{pixel_distributions} displays the probability densities for these distributions.
Notice that as the $\sigma$ increases, the distribution of the weights looks
more like those of a~$U(0,1)$,
and therefore, when doing binary classification, the misclassification rate
should increase as~$U(0,1)$ is compared to~$N(0.5,\sigma)$ with $\sigma$
increasing.
\figref{image-annulus} includes an example of an annulus with pixel intensities
drawn from a~$U(0,1)$
distribution and from a~$N(0.5, 0.17)$ distribution.

\paragraph{Function Extension.}
For these experiments, we consider the maximum and average extensions
(Equations~\eqref{eqn:max_extension} and \eqref{eqn:avg_extension}, respectively) for the weight functions to evaluate differences in classification performance.

\subsection{Classification Methods}

The seven shapes, four distributions, and two function extensions define 56 image classes.
The \wect is evaluated on the task of binary classification using support vector machines (SVM)~\cite{cortes1995support}
and $K$-nearest-neighbors \cite{knn1951}.
For each classification model, 250 images (half from each class) are independently generated,
with 80 percent of the images (200 images) used for training and the remaining 20 percent (50 images) used for testing.
To approximate the \wect, we sample equally-spaced directions
from $\sph^1$.
We discuss results below for choosing different numbers of directions from $\sph^1$.
To vectorize each \fwec,~$91$ equally-spaced threshold values from $-45$ to $45$ are considered.
This ensures that there is at least one discretization value for each height threshold in any direction
of filtration.

\paragraph{Implementation.}
The simulation study was implemented in Python, using the Dionysus~2 library~\cite{dionysus2}.
For classification tasks using SVM, we train an SVM separately for each binary classification model.
All SVMs in this work use a regularization parameter of value 20.
The classification was done using an SVM classifier library from Scikit-learn \cite{scikit-learn}.
For classification tasks using $K$-nearest-neighbors, a simple Python implmentation was created using
our implementation of the \wect distance function.

\subsection{Results}

\begin{figure}[tbh]
    \centering
    \includegraphics[height=1.5in]{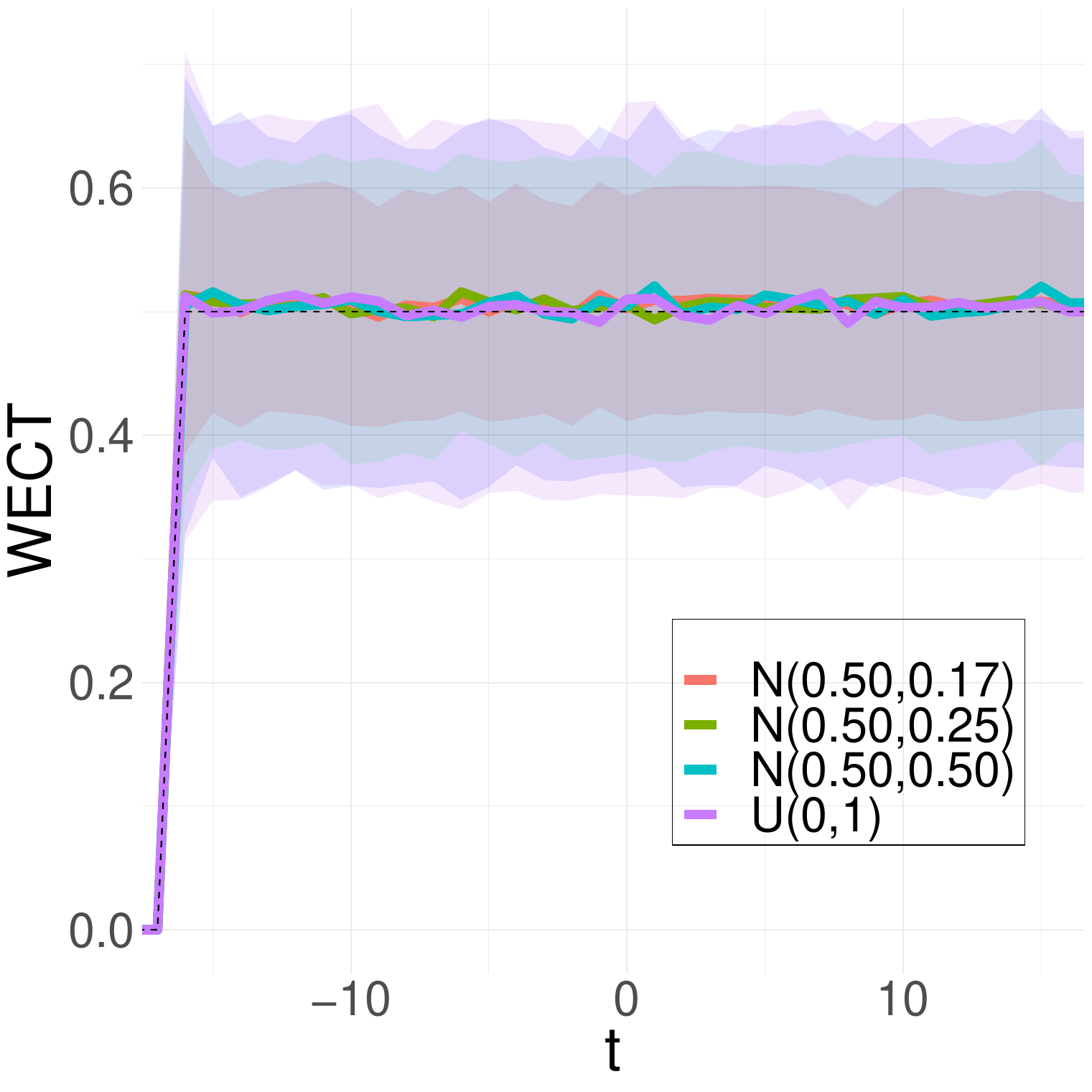}
    \caption{The \wect in one direction $s = \langle 1, 0 \rangle$ for four different probability distributions on a square using the $\Omega_{\text{avg}}$ extension.
    Each of the four distributions matches the expected \wect (the dotted line).
    Note that the entire vectorization occurs on the interval $[-45, 45]$, but for this direction the
    \wect is zero until height~$t=-17$ and remains constant after height~$t=17$. }
    \label{fig:square-expectation}
\end{figure}
To better understand the \wect, we considered several experiments
that provide insight into how different factors affect the \wect and
the information it encodes.
In particular, we analyze changes in the number of directions, pixel intensity distributions,
function extensions, and use of the \wect vs.\ the vectorized \wect.
However, first, as \figref{square-expectation} demonstrates, we verify that for all four pixel intensity
distributions, the computation of the \wect  matches the theoretical expectation
given in \eqnref{expWECTSquare}.

\begin{figure}
    \centering
    \includegraphics[width=\linewidth]{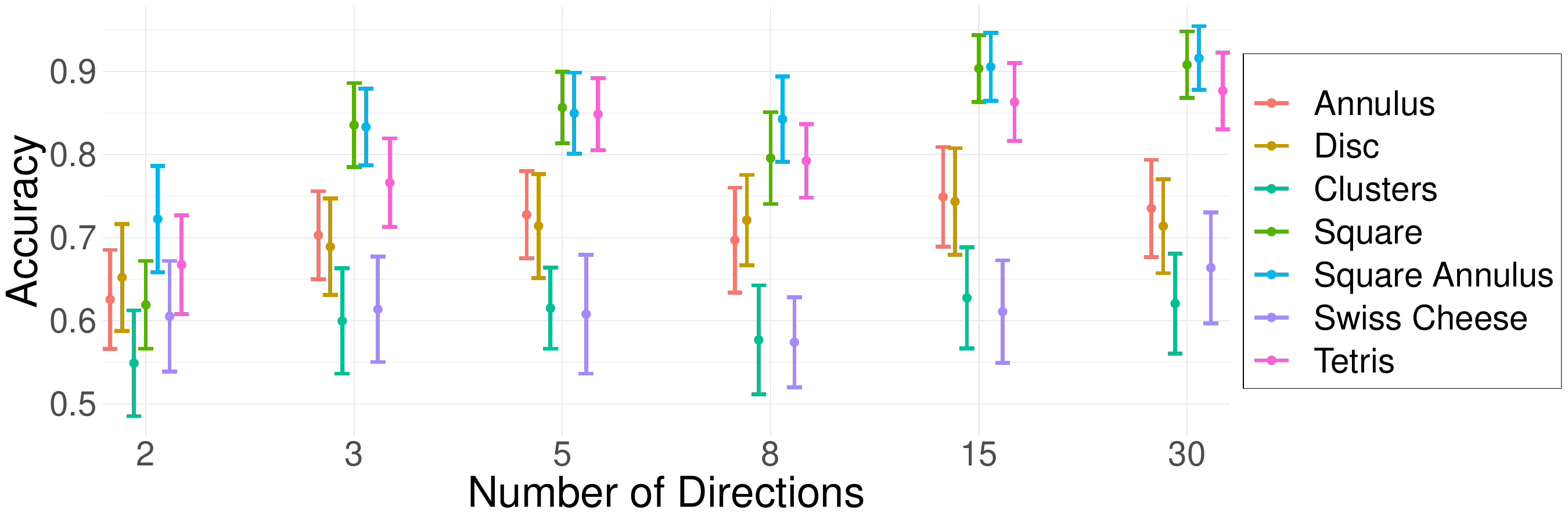}
    \caption{Classification accuracy of the $U(0,1)$ image against the $N(0.5,0.25)$ image for each shape across
    varying numbers of directions of filtration using SVM and the average function extension.
    As expected, increasing the number of directions had minimal effect on the classifier for
    shapes with no structural orientation (such as the Disc and Annulus).
    The Swiss Cheese and Clusters shapes also do not appear to be affected by changing the number
    of directions.
    However, the Square, Square Annulus, and Tetris all encoded more information as the number of directions increased. }
    \label{fig:changing-directions}
\end{figure}
Next, we analyze how changing the number of directions of the filtration impacts
the ability of the \wect to capture information about the underlying image.
In order to do so, the following classification task was executed using the SVM classifier.
For each shape, the~$U(0,1)$ image was compared against the
$N(0.5,25)$ in a binary classification task for a varying number of directions.
The average function extension was used when computing the \wects.
For a given shape, if the \wect successfully encoded the structure of the two different images, the classification
accuracy is closer to one.
The classifier accuracy was tested on all the shapes for $2,3,5,8,15$ and $30$ directions.
As \figref{changing-directions} shows, as the number of directions increases,
accuracy of the classifier improved only for some shapes.
In order to balance an increase of encoded information in the \wect with computational cost,
the remainder of the experiments are executed using 15 directions.

\begin{figure}
    \centering
    \begin{subfigure}{.48\textwidth}
        \centering
       \includegraphics[height=2.35in]{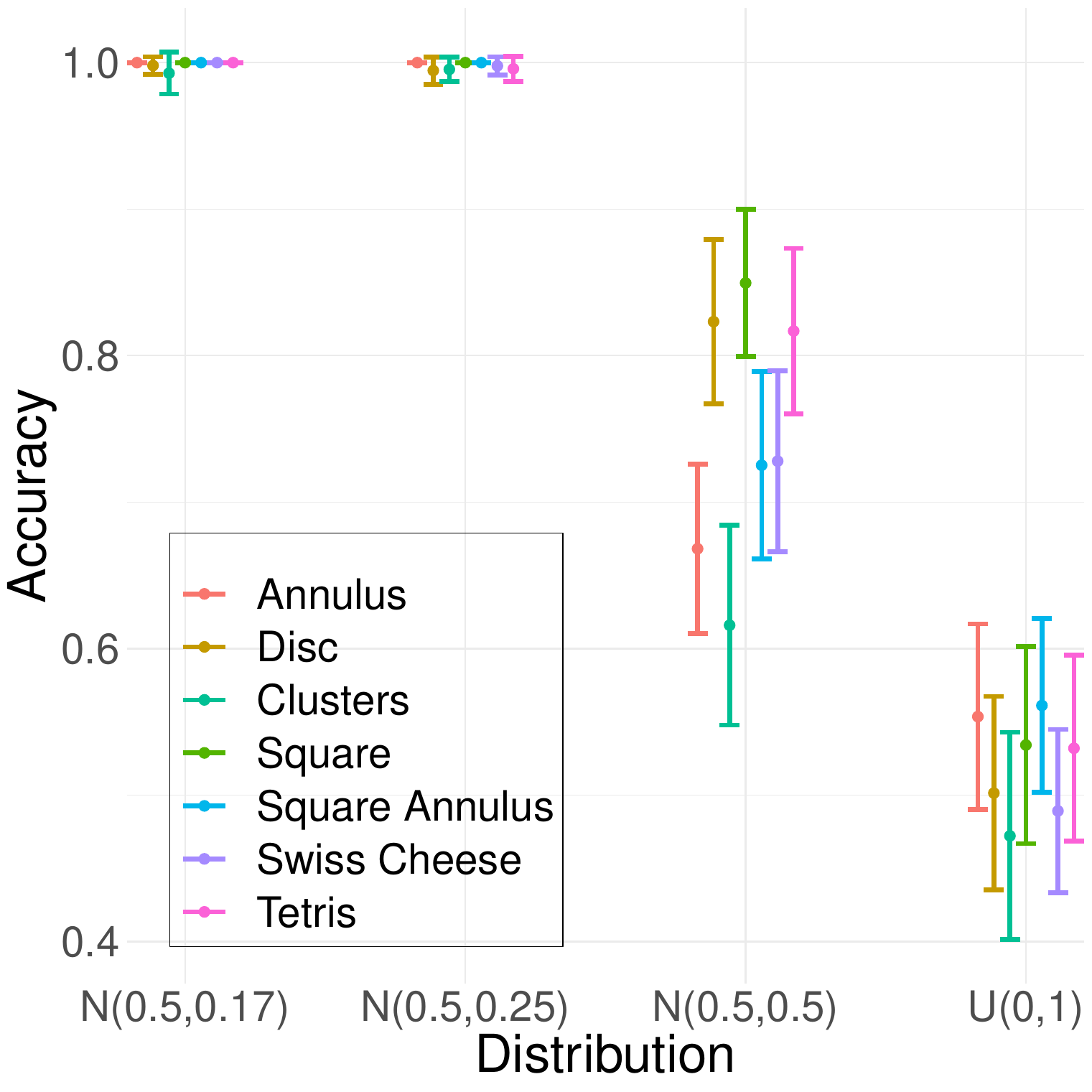}
        \caption{Maximum Extension}
        \label{fig:pairwiseResultsMaxFE}
    \end{subfigure}
    \hfill
    \begin{subfigure}{.48\textwidth}
        \centering
       \includegraphics[height=2.35in]{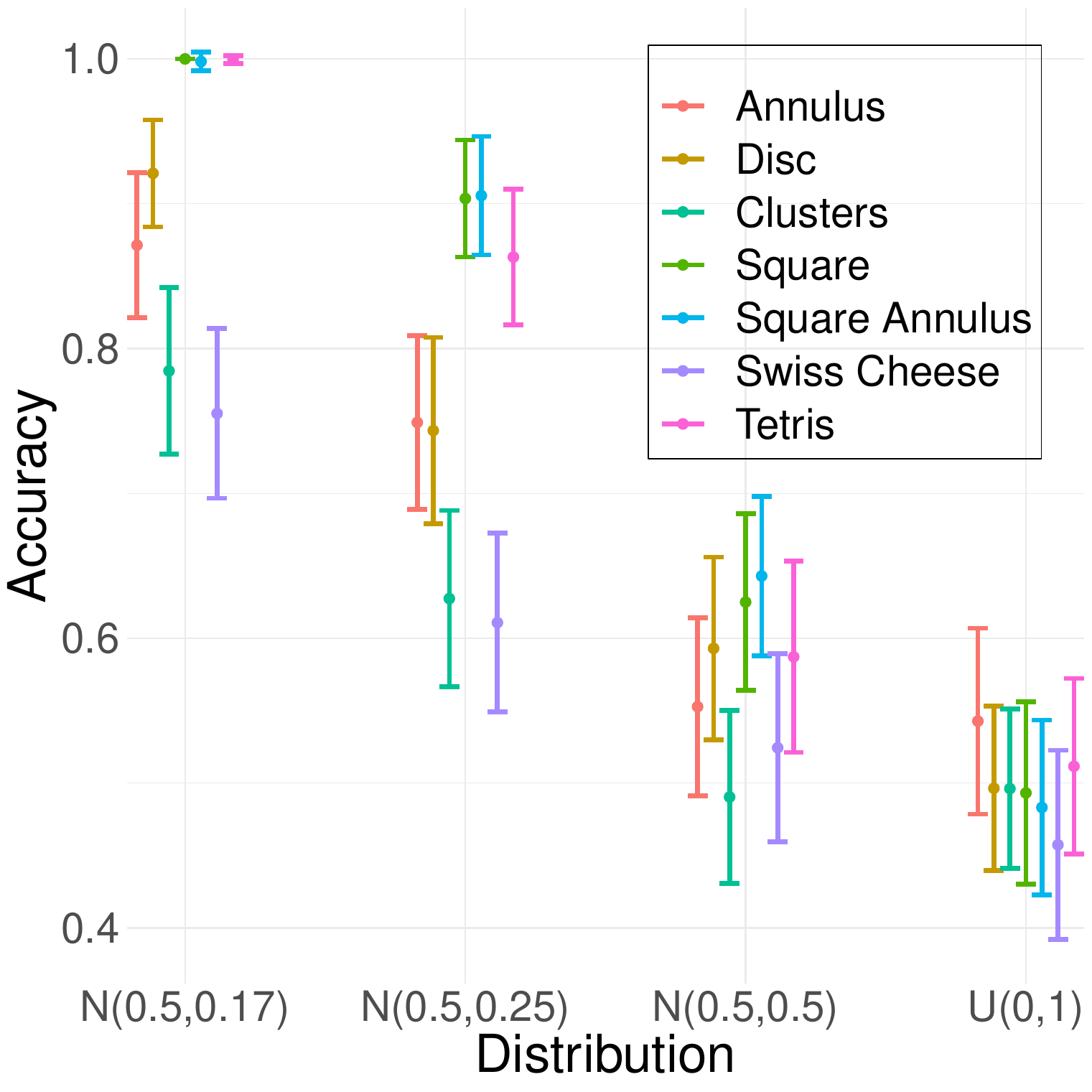}
        \caption{Average Extension}
        \label{fig:pairwiseResultsAvgFE}
    \end{subfigure}
    \caption{Binary testing classification accuracy (percentage of correct
    classifications out of the 50 test images)  using a SVM of the classes
    indicated by the pixel intensity distribution (x-axis) and image shape
    (color) versus the same image shape with $U(0,1)$ pixel intensities.
    The~$U(0,1)$ on the x-axis is a control setting where the accuracy should be
    around $0.5$ since the two classes considered in the model have the same
    image shape and the same pixel intensity distribution.  Note that the other
    combinations of classes are excluded here because the accuracy was nearly
    perfect (greater than $0.99$) regardless of the pixel intensity
    distribution.}
    \label{fig:pairwiseResults}
\end{figure}
In order to gain a deeper understanding of the \wect properties, we conducted experiments to observe how changing pixel
intensity distributions in various filtration directions affect its behavior.
Again, we examined the \wect's capacity to extract image information through binary
classification tasks using an SVM classifier.
For each shape, we consider the task of binary classification of \wects in $15$ directions from each of the four distributions
against the \wect from the $U(0,1)$ distribution.
All pairwise classification models were considered, but when image shape differed the
classifier performed nearly perfect regardless of the pixel intensity distribution so those results are not summarized further.
Results where the two classes have the same shape can be found in \figref{pairwiseResults} for both function extensions.
As anticipated, the accuracy of the classifier for~$U(0,1)$ against~$U(0,1)$ is
roughly random,~$U(0,1)$ against~$N(0.5, 0.17)$ is nearly perfect,~$U(0,1)$ against $N(0.5, 0.25)$ is slightly worse,
and~$U(0,1)$ against $N(0.5, 0.5)$ is barely better than random.
Ultimately, the decrease in accuracy as the normal distribution becomes more similar to the uniform distribution (i.e., as $\sigma$ increases)
demonstrates that the \wect encodes information about the pixel intensity distribution.
In other words, \emph{the \wect captures and encodes structural information from the images that is not encoded by the ECT.}

Next, we compare the results using the maximum function extension (\figref{pairwiseResultsMaxFE})
and the average function extension (\figref{pairwiseResultsAvgFE}).
Overall, across nearly all the distributions, the classifier performs better
on \wects computed using the maximum function extension as opposed to the average function extension.
As the distributions differ more at their extremes in the current setting, it is likely that
the maximum function extension better emphasizes these differences, which in turn increases the
accuracy of the~classifier.

\begin{figure}
    \centering
    \begin{subfigure}{.49\textwidth}
        \centering
        \includegraphics[height=1.95in]{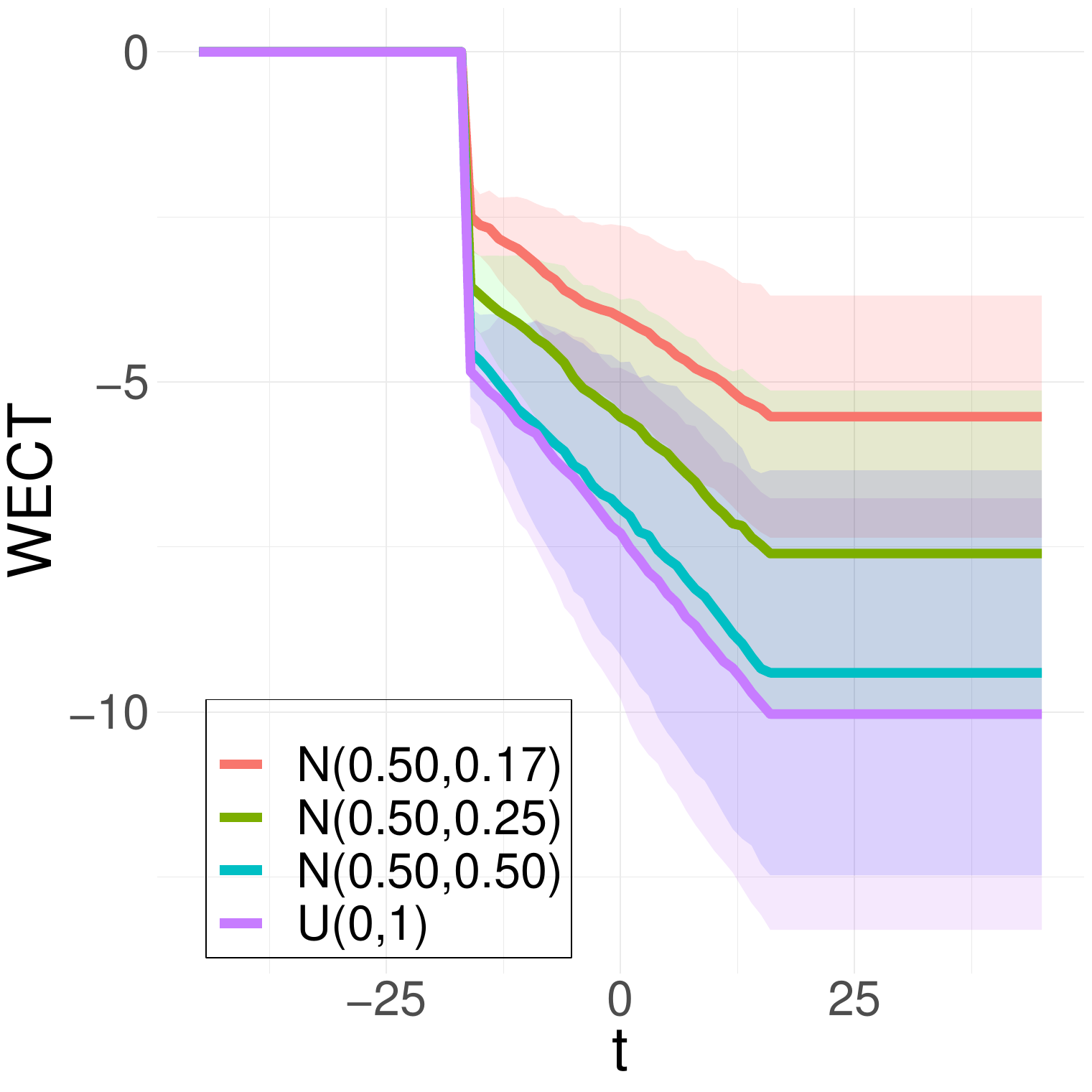}
        \caption{Maximum Extension}
        \label{fig:maxsquare_mean}
    \end{subfigure}
    \begin{subfigure}{.49\textwidth}
        \centering
        \includegraphics[height=1.95in]{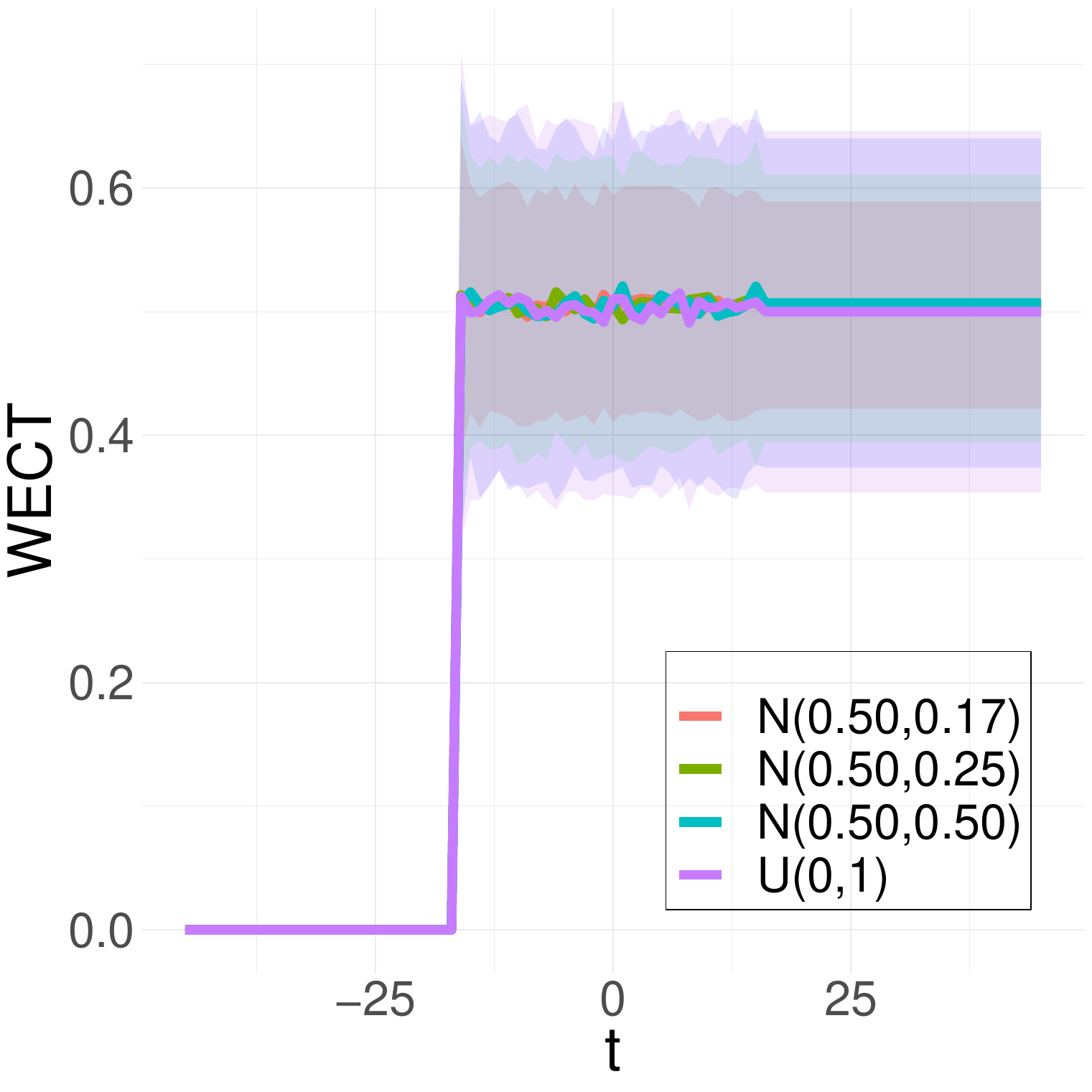}
        \caption{Average Extension}
        \label{fig:avgsquare_mean}
    \end{subfigure}
    \caption{The average \wect for one direction $s = \langle 1, 0 \rangle$ of the 250 generated images of the square.
    The colored bands indicate the region within one standard deviation of the mean.
    Observe that the \wect changes at $t=-17$, where the height filtration passes the left side of the square,
    and stops changing at $t=17$, where the height filtration passes the right side of the square.
    The \wect for all distributions follow a similar pattern.
    For the maximum extension, as the normal distribution becomes more
    uniform, the corresponding \wect approaches the \wect from the uniform distribution.
    }
    \label{fig:square_avgwect}
\end{figure}
\begin{figure}
    \centering
    \begin{subfigure}{.31\textwidth}
        \centering
       \includegraphics[height=1.5in]{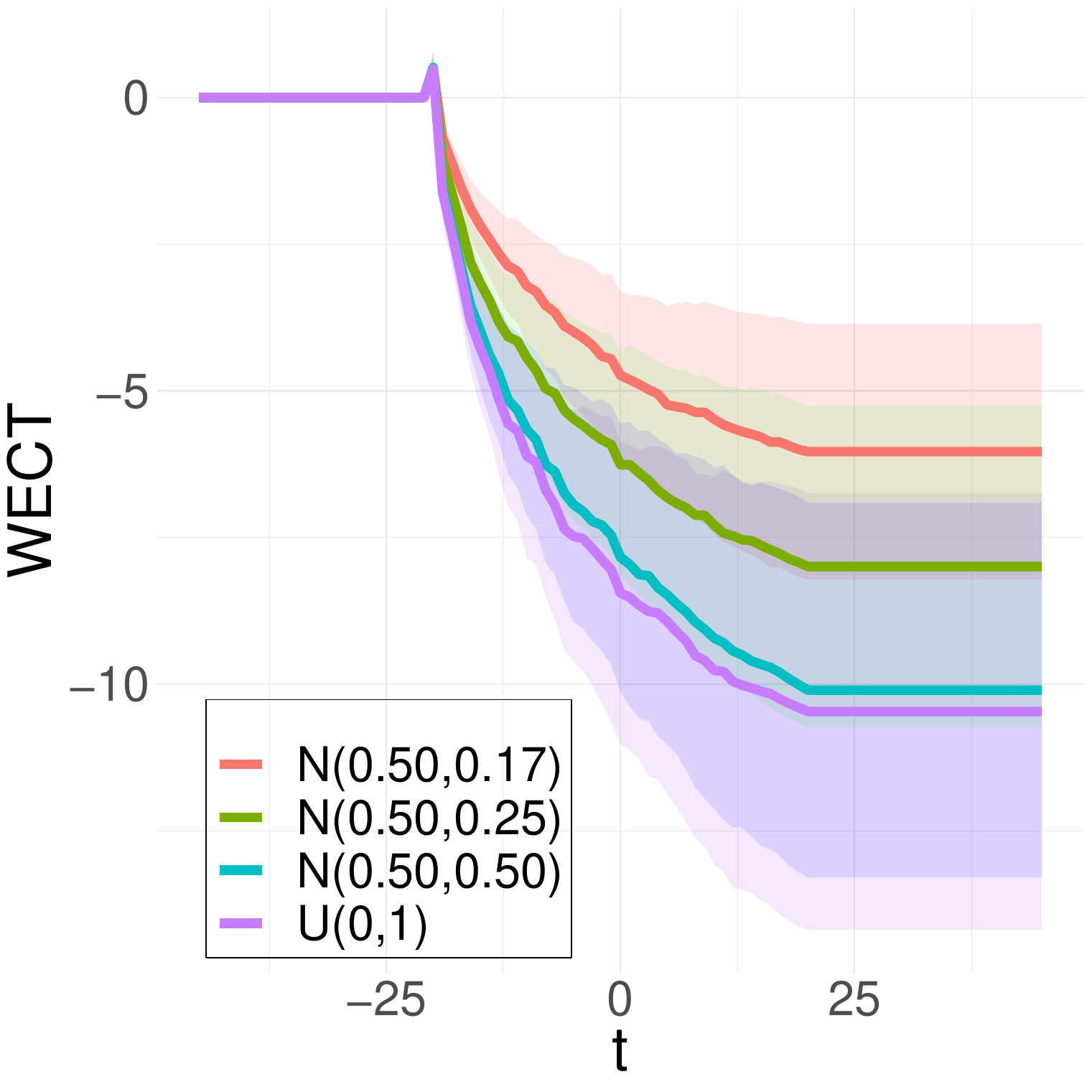}
        \caption{Disc}
        \label{fig:maxdisc_mean}
    \end{subfigure}
    \hfill
    \begin{subfigure}{.31\textwidth}
        \centering
       \includegraphics[height=1.5in]{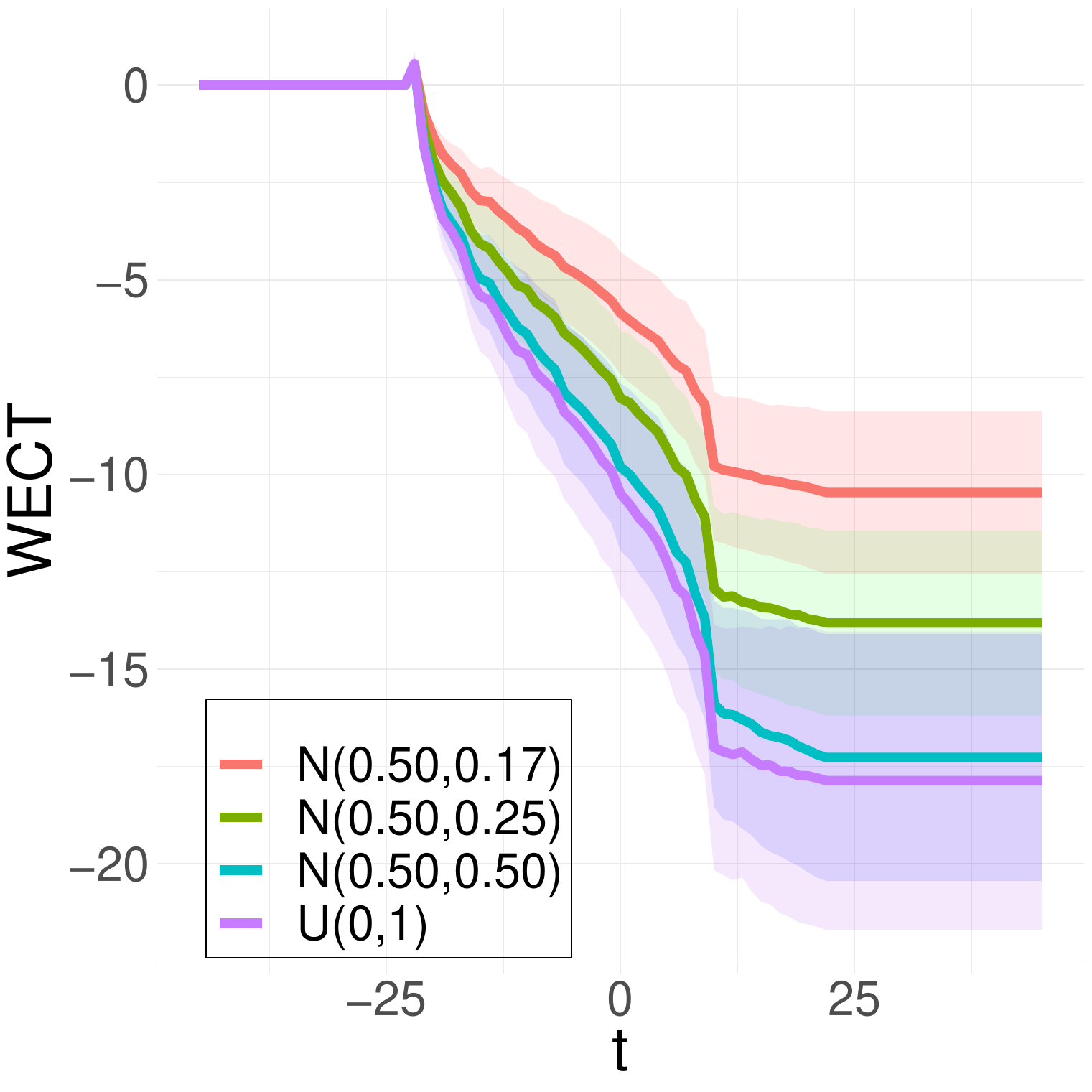}
       \caption{Annulus}
        \label{fig:maxannulus_mean}
    \end{subfigure}
    \hfill
    \begin{subfigure}{.31\textwidth}
        \centering
       \includegraphics[height=1.5in]{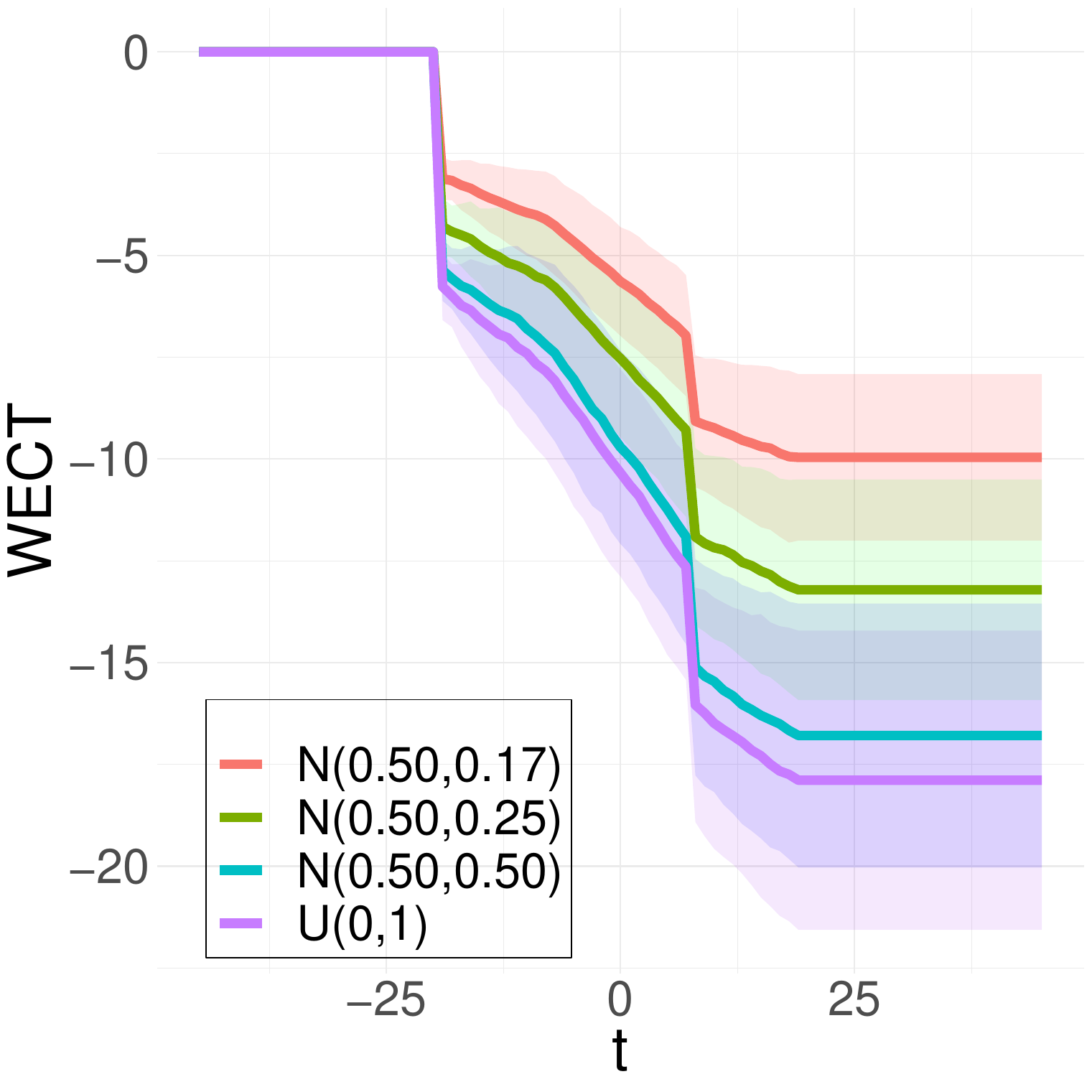}
       \caption{Square Annulus}
        \label{fig:maxsquare_annulus_mean}
    \end{subfigure}
    \\~\\
    \begin{subfigure}{.31\textwidth}
        \centering
       \includegraphics[height=1.5in]{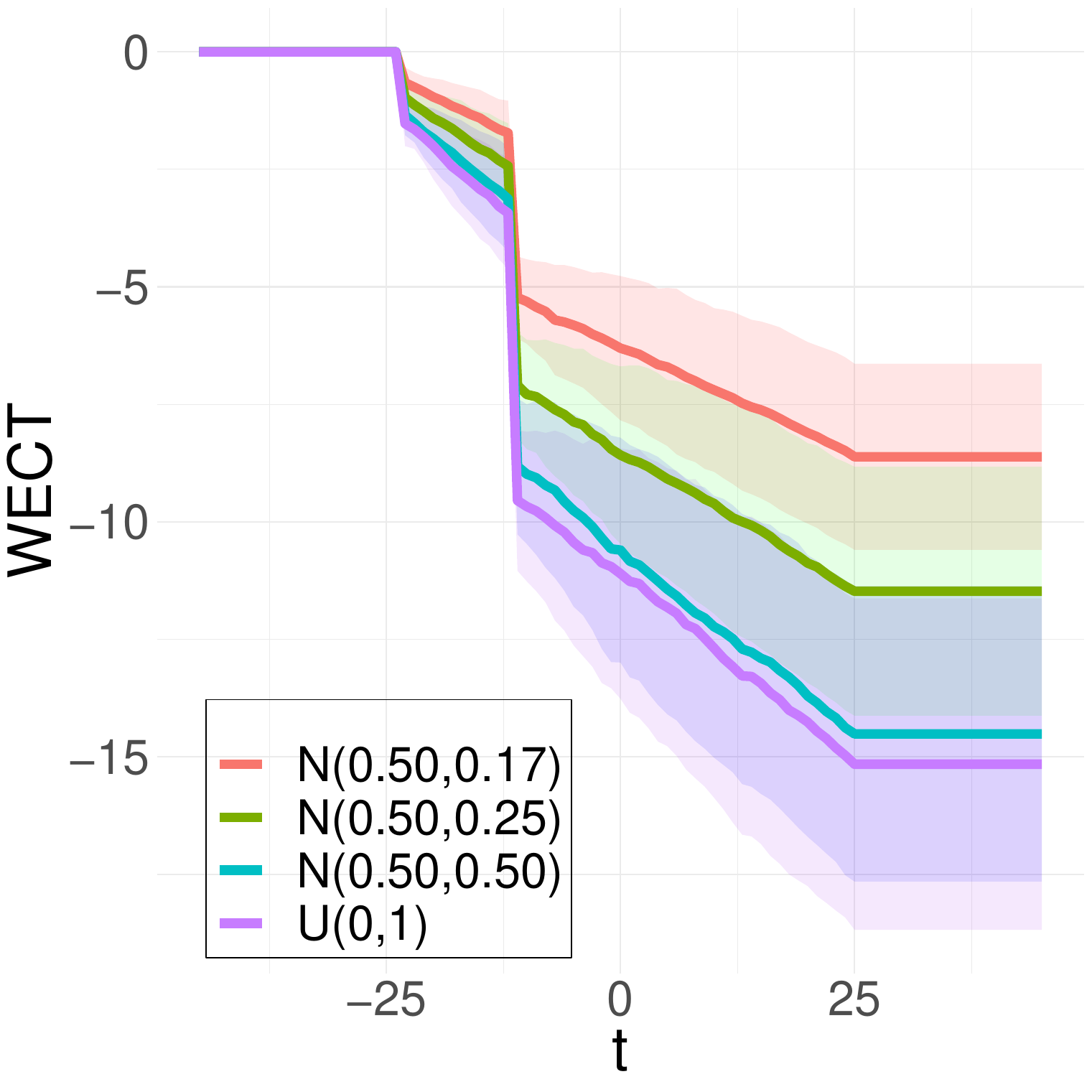}
        \caption{Tetris}
        \label{fig:maxtetris_mean}
    \end{subfigure}
    \hfill
    \begin{subfigure}{.31\textwidth}
        \centering
       \includegraphics[height=1.5in]{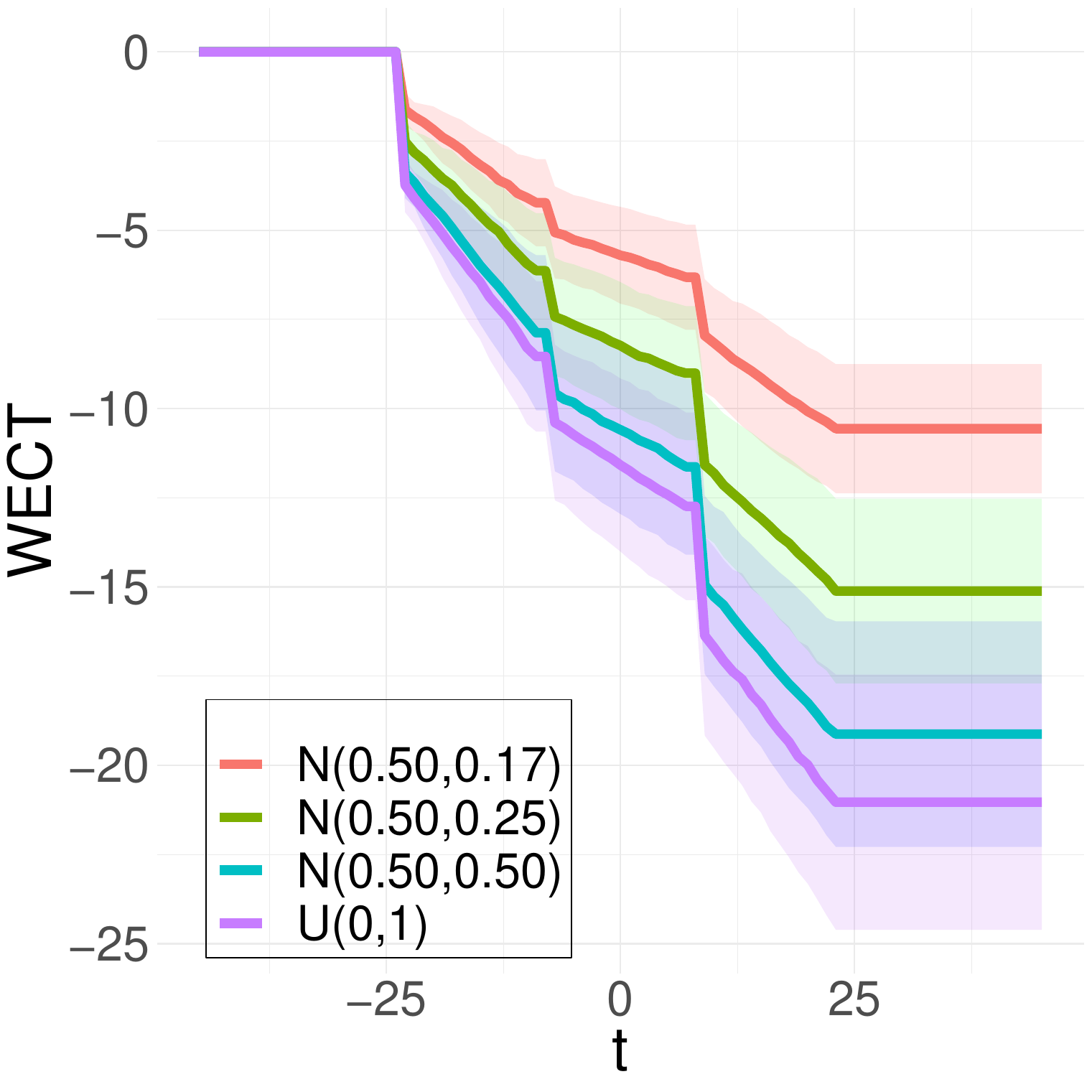}
        \caption{Clusters}
        \label{fig:maxclusters_mean}
    \end{subfigure}
    \hfill
    \begin{subfigure}{.31\textwidth}
        \centering
       \includegraphics[height=1.5in]{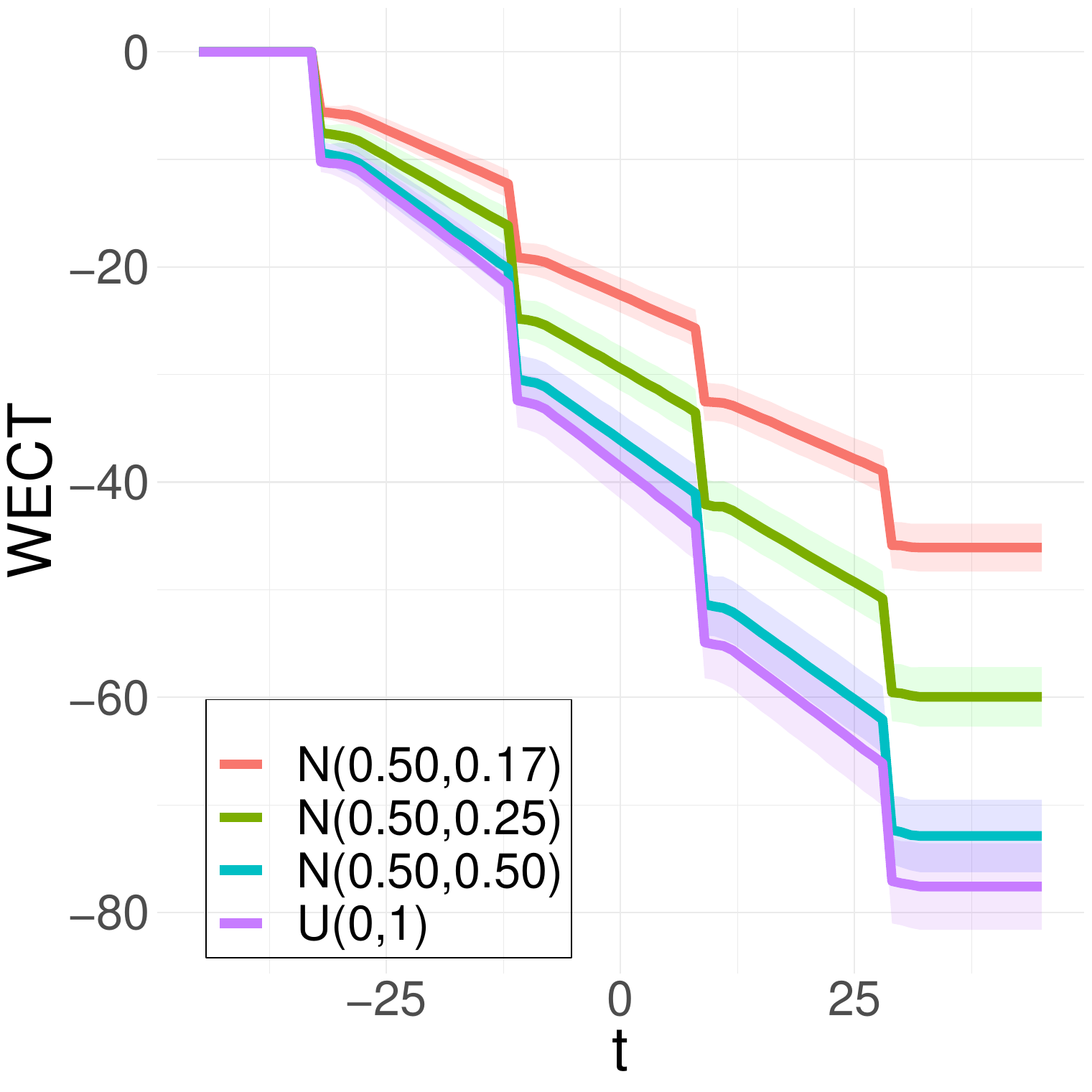}
        \caption{Swiss cheese}
        \label{fig:maxswiss_mean}
    \end{subfigure}
    \caption{The average \wect for one direction of the 250 images generated using the maximum extension.
    Note that the y-axis scales differ among the plots to better observe small-scale features of the functions.
    The \wects corresponding to the normal distributions approach the \wect corresponding to the uniform distribution
    as the normal distribution becomes closer to uniform.
    Additionally, similar shapes have similar patterns, such as the two versions of the annulus.}
    \label{fig:maxwect_mean}
\end{figure}
\begin{figure}
    \centering
    \begin{subfigure}{.31\textwidth}
        \centering
       \includegraphics[height=1.5in]{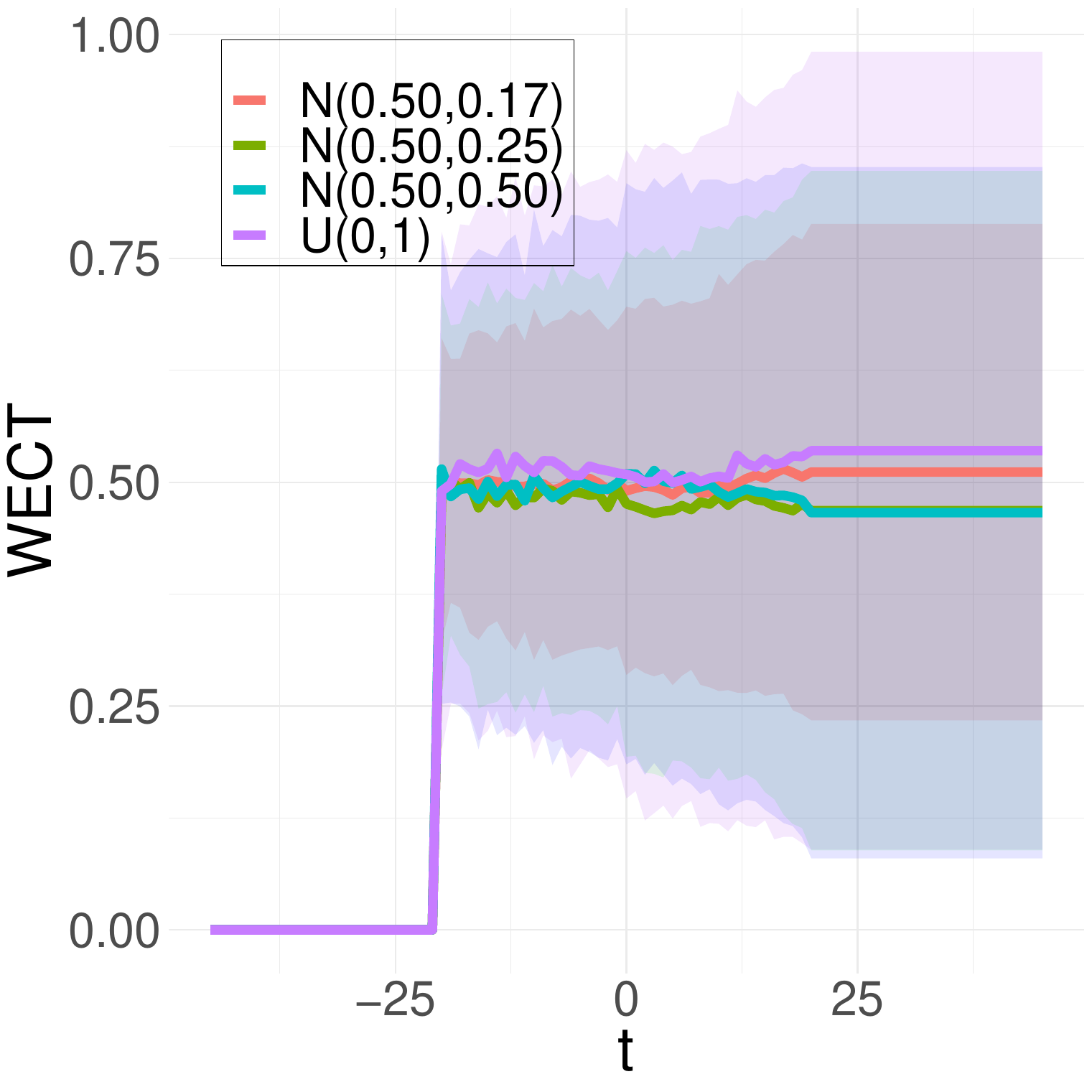}
        \caption{Disc}
        \label{fig:avgdisc_mean}
    \end{subfigure}
    \hfill
    \begin{subfigure}{.31\textwidth}
        \centering
       \includegraphics[height=1.5in]{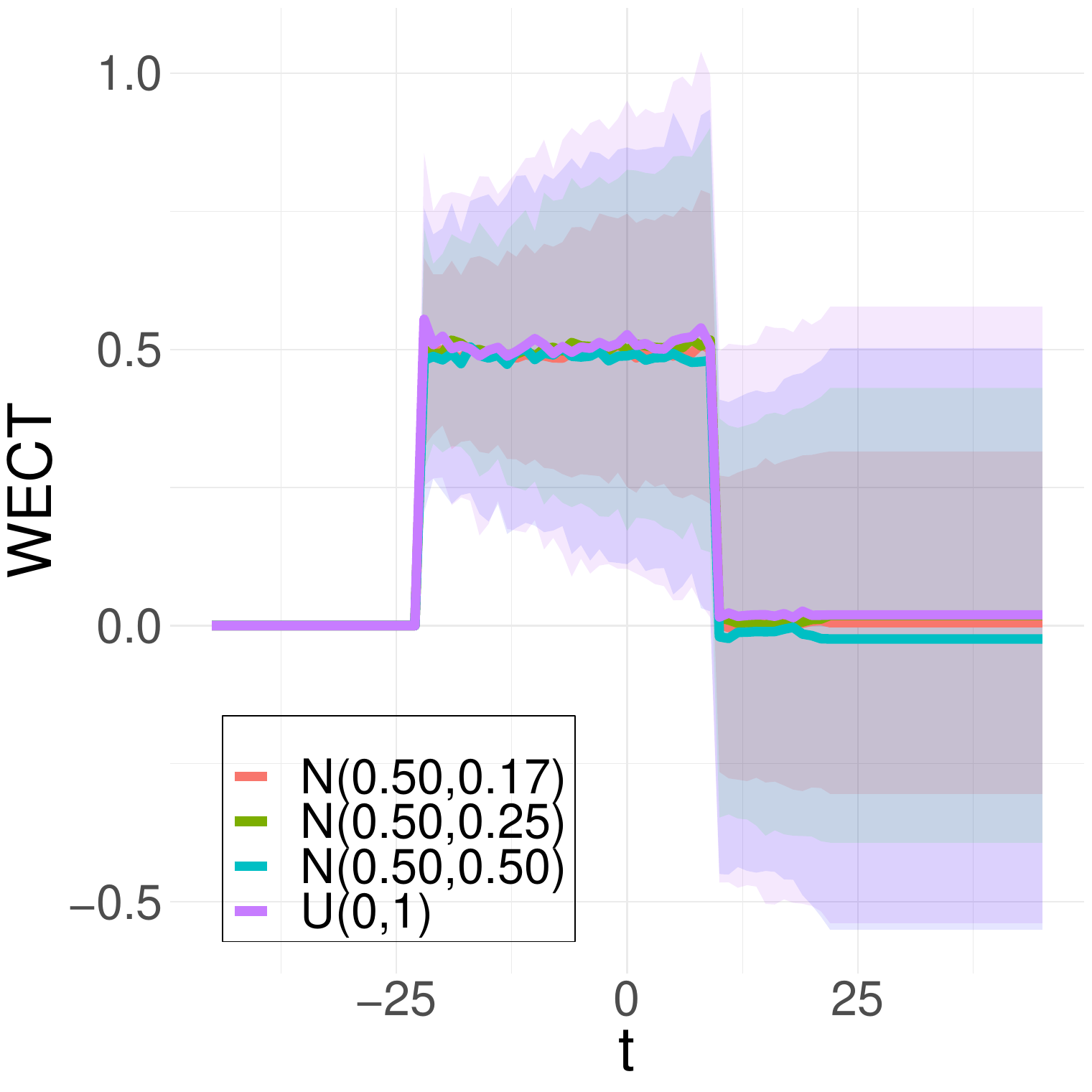}
       \caption{Annulus}
        \label{fig:avgannulus_mean}
    \end{subfigure}
    \hfill
    \begin{subfigure}{.31\textwidth}
        \centering
       \includegraphics[height=1.5in]{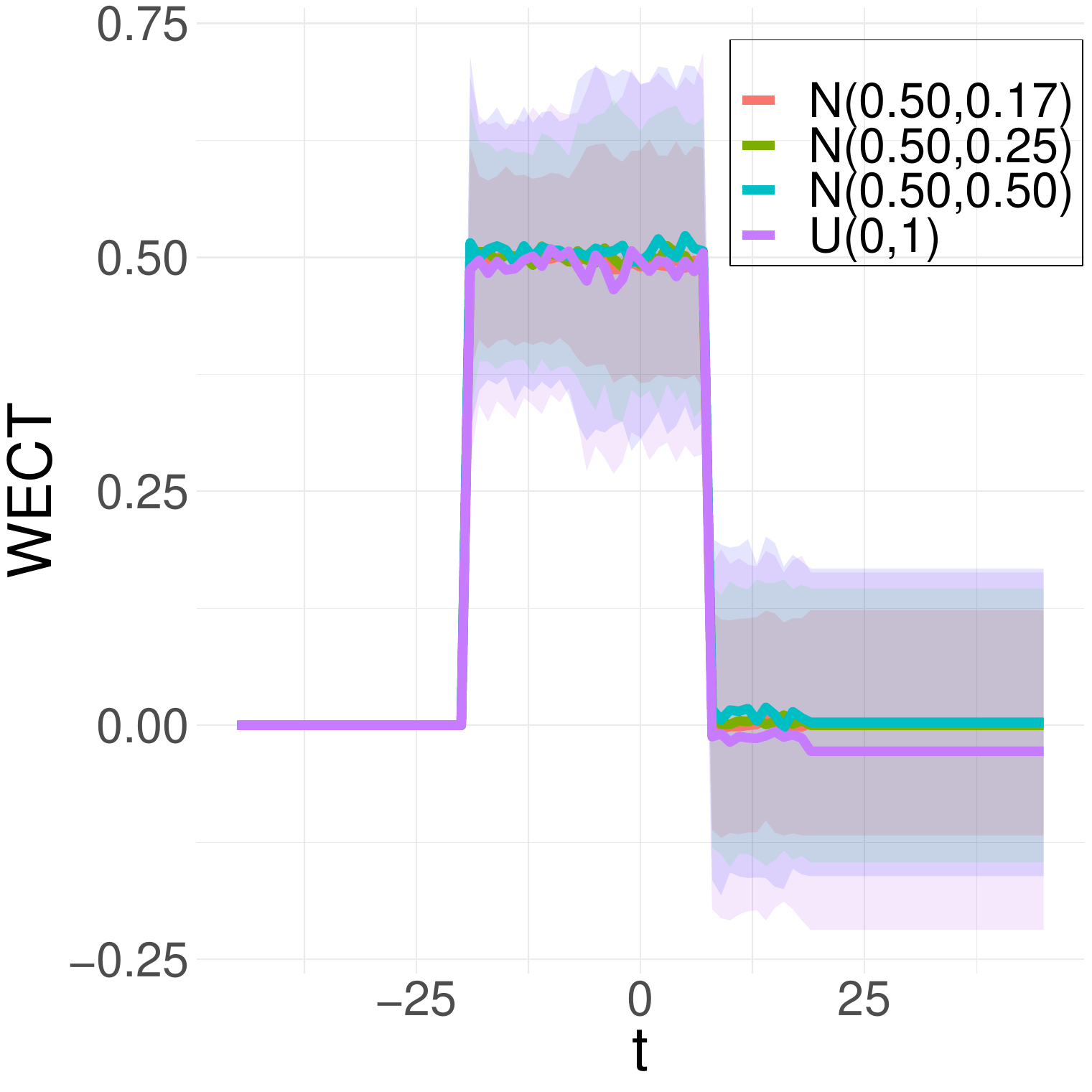}
       \caption{Square Annulus}
        \label{fig:avgsquare_annulus_mean}
    \end{subfigure}
    \\~\\
    \begin{subfigure}{.31\textwidth}
        \centering
       \includegraphics[height=1.5in]{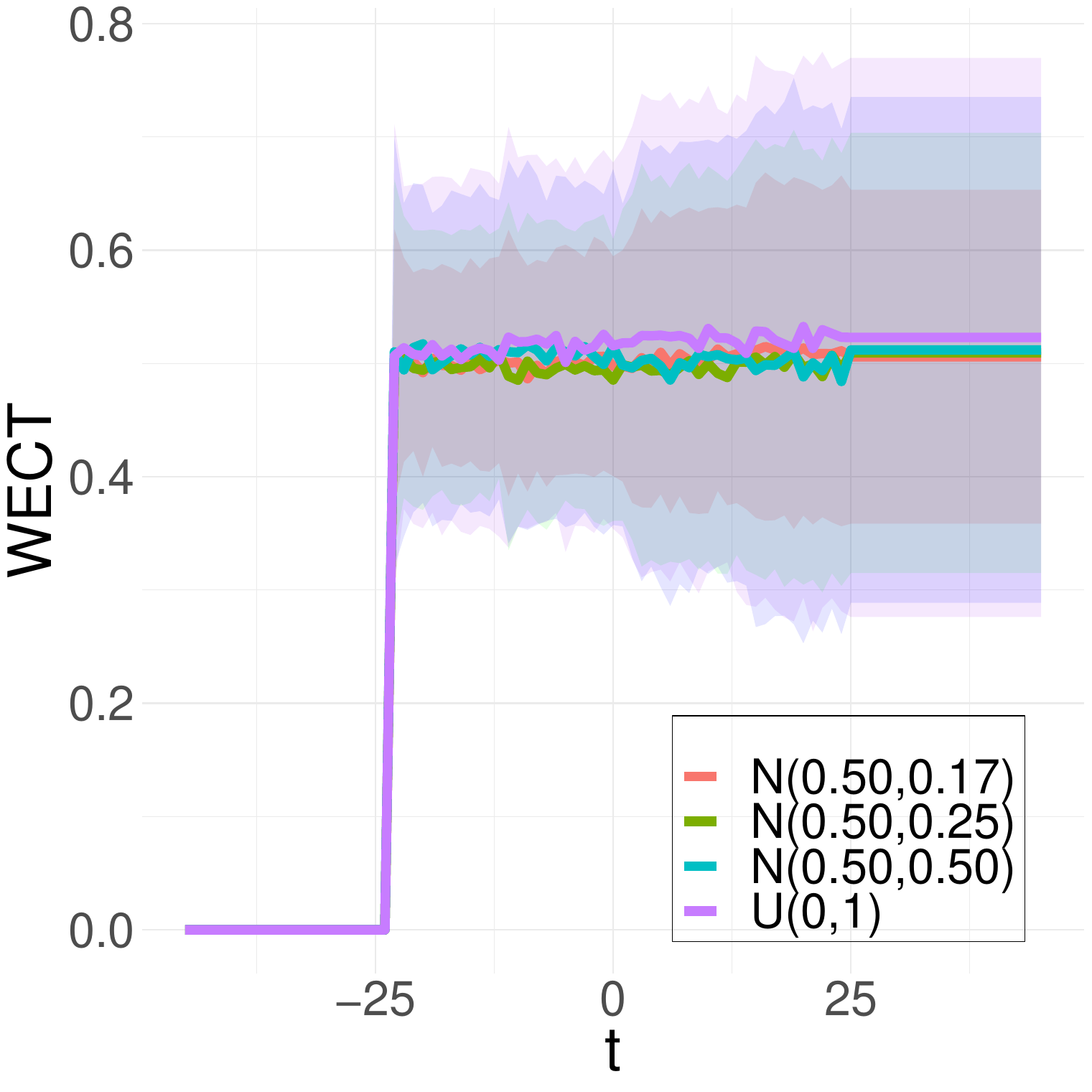}
        \caption{Tetris}
        \label{fig:avgtetris_mean}
    \end{subfigure}
    \hfill
    \begin{subfigure}{.31\textwidth}
        \centering
       \includegraphics[height=1.5in]{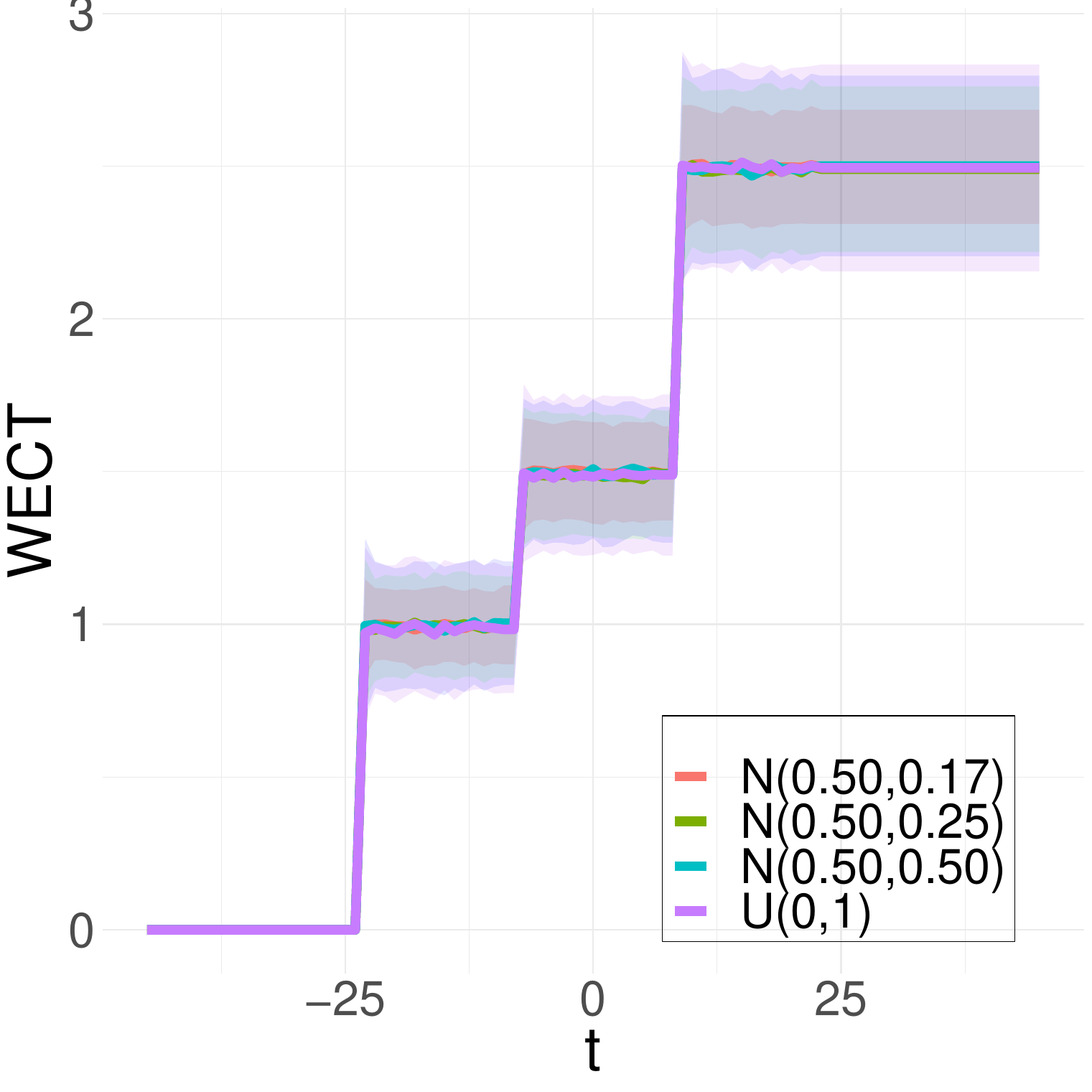}
        \caption{Clusters}
        \label{fig:avgclusters_mean}
    \end{subfigure}
    \hfill
    \begin{subfigure}{.31\textwidth}
        \centering
       \includegraphics[height=1.5in]{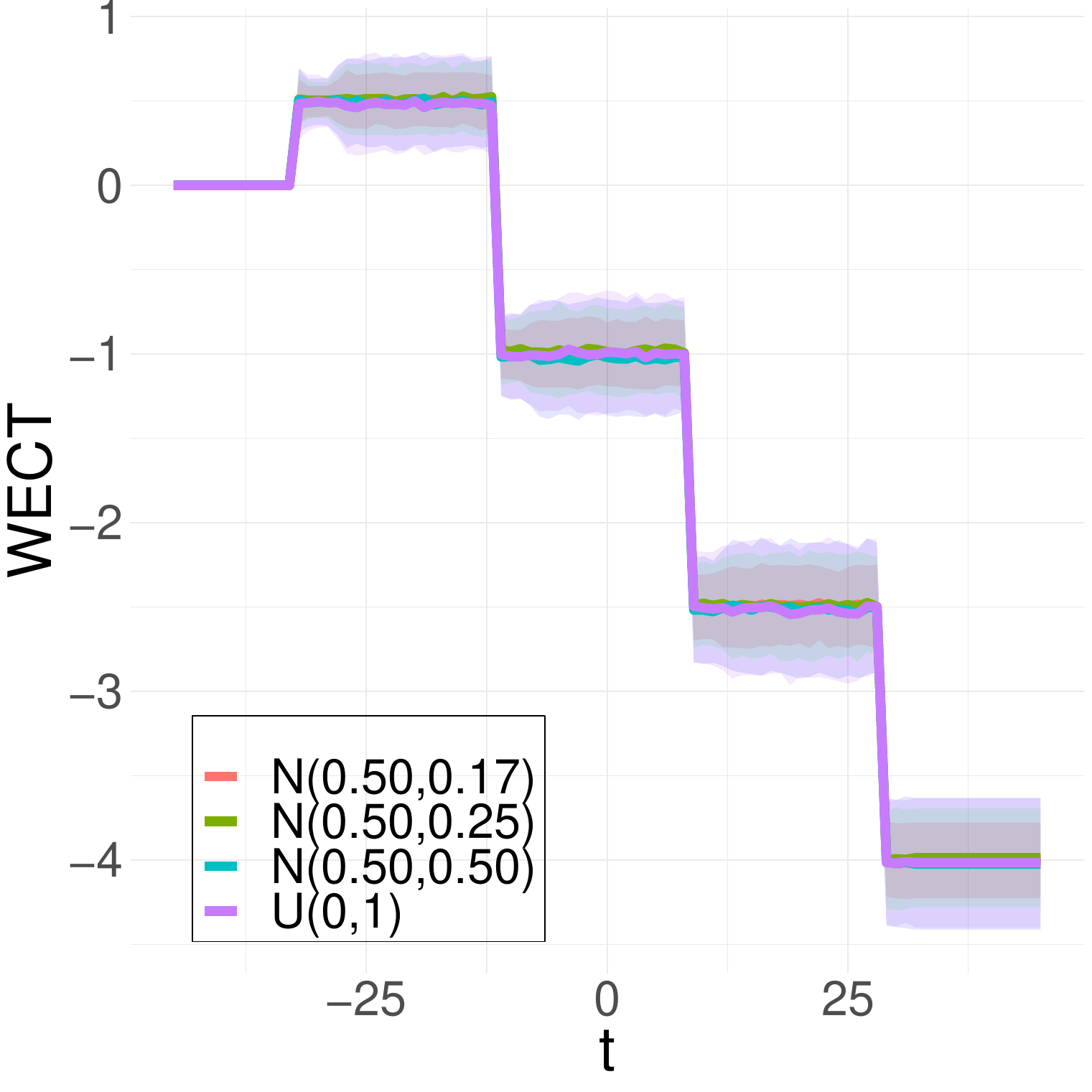}
        \caption{Swiss cheese}
        \label{fig:avgxswiss_mean}
    \end{subfigure}
    \caption{The average \wect for one direction of the $250$ images generated using the average extension.
    Observe similarities in similar shapes, like the two versions of the annulus.
    Additionally, the symmetry between the Clusters and the Swiss cheese shapes
    is evident in the \wects.
    Note that y-axis scales~differ.}
    \label{fig:avgwect_mean}
\end{figure}

To better understand the classification performance in the above experiments,
we consider visualizations of the \wect for the setting that considers fifteen directions.
\figref{square_avgwect} displays the average \wect
in one direction of 250 generated images under the two different function extensions.
The shaded region indicates the values within one standard deviation of the mean.
For both function extensions, the average \wect generally follows the same pattern,
but the values of the \wect depend on the distribution of the pixel intensities.
The \wect using the $U(0,1)$ distribution generally has the lowest values,
and the \wect using the $N(0.50, 0.50)$ and distribution tends to have slightly greater values, then
$N(0.50, 0.25)$, and finally the $N(0.50, 0.17)$ distribution tends to have the highest values.
These results hold also for the rest of the shapes, as indicated in \figref{maxwect_mean} and \figref{avgwect_mean}.
This, along with the classification results, indicate that incorporating the weights into the ECT is important for
distinguishing images generated under different intensity distributions.
Note also that the average \wect for different shape classes have varying levels and patterns of smoothness, which reveals why the
classification rate between different shapes is nearly perfect.

\begin{figure}
    \centering
    \includegraphics[height=3in]{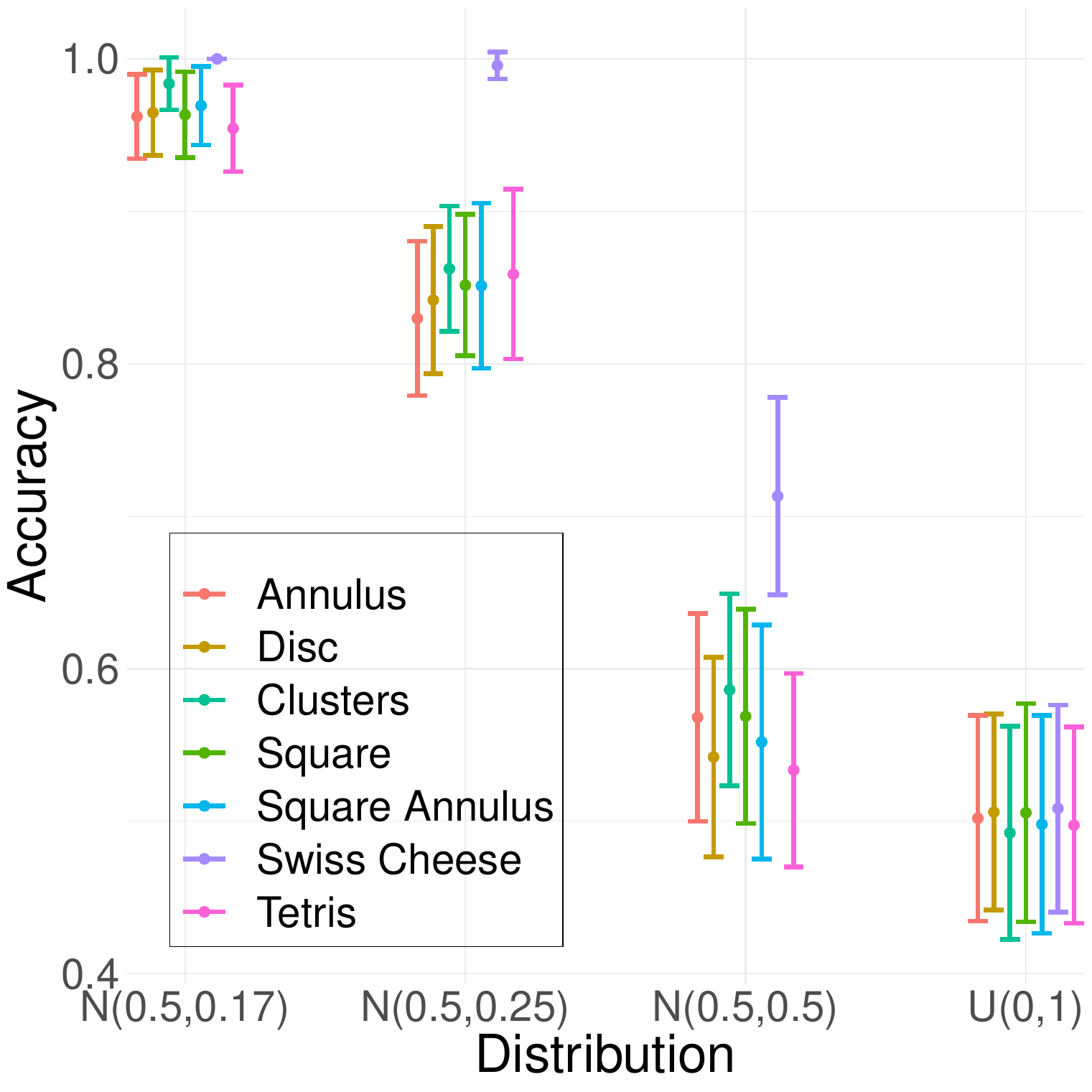}
    \caption{Binary classification of the four distributions
    against the~$U(0,1)$ distribution for each shape using KNN.
    The \wects were computed in eight directions using the maximum function extension.
    Note that experiments were restricted to the eight direction case due to the high computational cost.
    The trend matches that of SVM using the vectorized \wects in previous
    experiments, indicating that the \wect distance and
    vectorized \wect encode similar information about the underlying images.
    }
    \label{fig:knnsvm}
\end{figure}
A final result is the use of the
true (non-vectorized) \wect in the classification experiments.
As explained in Section~\ref{par:distances}, one can compute a distance between
two \wects by integrating the difference between the piecewise \wecfs in each direction
(and taking the maximum difference across all directions).
Thus, a natural extension use of this distance is in a nonparametric classification
method such as $K$-nearest-neighbors (KNN).
\figref{knnsvm} summarizes results for the KNN classifier
based on the \wect distance with $K=5$ for eight directions with the maximum extension.
In general, it appears that performance of SVM was better in these binary classification tasks.
Extensive experiments with KNN were less feasible due to the high computational
cost of nonparametric methods.
However, the focus is not on using the \wect as a classifier, but instead
using the classification tasks to understand the information that is encoded in
the \wect.
For both methods, we observe a decrease in accuracy as the pixel intensity distribution becomes more similar,
which indicates that both the vectorized \wect and the \wect distance successfully encode
underlying information about the image.

% \begin{figure}
%     \centering
%     \begin{subfigure}{.48\textwidth}
%         \centering
%        \includegraphics[height=2.35in]{figs/results/knn_svm_n17.pdf}
%         \caption{$N(0.5,0.17)$}
%         \label{fig:knnsvm_n17}
%     \end{subfigure}
%     \hfill
%     \begin{subfigure}{.48\textwidth}
%         \centering
%        \includegraphics[height=2.35in]{figs/results/knn_svm_n25.pdf}
%         \caption{$N(0.5,0.25)$}
%         \label{fig:knnsvm_n25}
%     \end{subfigure}
%     \\~\\
%     \begin{subfigure}{.48\textwidth}
%         \centering
%        \includegraphics[height=2.35in]{figs/results/knn_svm_n50.pdf}
%         \caption{$N(0.5,0.5)$}
%         \label{fig:knnsvm_n50}
%     \end{subfigure}
%     \hfill
%     \begin{subfigure}{.48\textwidth}
%         \centering
%        \includegraphics[height=2.35in]{figs/results/knn_svm_uniform.pdf}
%         \caption{$U(0,1)$}
%         \label{fig:knnsvm_uniform}
%     \end{subfigure}
%     \caption{\todo{}}
%     \label{fig:knnsvm}
% \end{figure}

\section{Conclusion}\label{sec:discussion}
While the ECT has been successful at summarizing shapes in a variety of contexts, image data poses a unique challenge because the pixel intensities potentially carry significant and relevant information.  The WECT was proposed to generalize the ECT and allow for the incorporation of pixel intensities as weights.  However, the intuition and interpretation of the WECT is not as clear.
In this paper, we explored using the \wect to represent shapes found in images, and developed an understanding of the importance of the weights.  Indeed, we found that the \wect captured more than the EC and
the ECF alone.  
This was especially important when images have the same shape, but different intensity distributions as assessed in our empirical study.  As the intensity distributions become more similar between two classes that have the same shape, the test classification accuracy decreases.
When the two classes had different image shapes, the \wect correctly classified new images perfectly (or almost perfectly with test classification accuracy greater than 0.99).  While we limited our classification models to SVM, and explored KNN models, \wects can be used in other classification models.

The effect on changing inputs to the \wect was also explored.  Two function
extensions, the maximum and average function extensions, were considered.
Though the maximum function extension outperformed the average function
extension in our settings, this was likely dependent on the pixel intensity
distributions considered.  In particular, the four distributions ($U(0,1)$,
$N(0.5,0.17)$,~$N(0.5,0.5)$, and~$N(0.5,5)$; all truncated to $(0,1]$) had
averages of $0.5$ so differences were near the extremes (near zero or one).
Therefore, the maximum function extension would amplify these differences better
than the~average.  

We also explored how the test classification accuracy changed with different
numbers of directions considered.  Shapes with rotational symmetry did not
benefit from increasing the number of directions, while those without rotational
symmetry did benefit.  How to appropriately select the number of directions in
this setting is an open question that will be explored in future~work.

Though classification was explored in this paper, the \wect could also be used for inference tasks such as hypothesis testing.  In light of this, we developed equations for the expected weighted EC and expected WECT for images, which may be useful for developing test statistics in future studies.
While we have found a clear benefit in using the \wect for image classification when the pixel intensity distributions differ, questions remain regarding its advantage for inference or for different data types.

% \camera{additional todos:
% * check for passive voice
% * site YRF version?
% }

% BTF: I am commenting this out, as grant ack. are above. I don't think we have
% additional aks
% \begin{acknowledgements}
%     \todo{}
% \end{acknowledgements}

% BibTeX users please use one of
% \bibliographystyle{spbasic}      % basic style, author-year citations
\bibliographystyle{siam}      % mathematics and physical sciences
% \bibliographystyle{spphys}       % APS-like style for physics
%\bibliography{}   % name your BibTeX data base
\bibliography{lama}

% Non-BibTeX users please use
% \begin{thebibliography}{}
% %
% % and use \bibitem to create references. Consult the Instructions
% % for authors for reference list style.
% %
% \bibitem{RefJ}
% % Format for Journal Reference
% Author, Article title, Journal, Volume, page numbers (year)
% % Format for books
% \bibitem{RefB}
% Author, Book title, page numbers. Publisher, place (year)
% % etc
% \end{thebibliography}

\end{document}